\documentclass[twocolumn,english,amssymb,aps,prl,superscriptaddress,tightenlines,notitlepage,longbibliography]{revtex4-2}

\usepackage{lmodern}
\usepackage[T1]{fontenc}
\usepackage[latin9]{inputenc}
\usepackage{color}
\usepackage{babel}
\usepackage{verbatim}
\usepackage{amsmath}
\usepackage{amssymb}
\usepackage{cancel}
\usepackage{graphicx}
\usepackage{esint}
\usepackage[unicode=true,pdfusetitle,
 bookmarks=true,bookmarksnumbered=true,bookmarksopen=false,
 breaklinks=false,pdfborder={0 0 1},backref=false,colorlinks=true]
 {hyperref}
\hypersetup{
 linkcolor=magenta,urlcolor=blue,citecolor=blue,pdfstartview={FitH},hyperfootnotes=false,bookmarks=False}

\makeatletter
\usepackage{babel}
\usepackage{xcolor}
\usepackage{graphicx}
\allowdisplaybreaks
\renewcommand{\big}{\bBigg@\@ne}
\renewcommand{\Big}{\bBigg@{1.5}}
\renewcommand{\bigg}{\bBigg@\tw@}
\renewcommand{\Bigg}{\bBigg@{2.5}}

\newcommand{\biggg}{\bBigg@\thr@@}
\newcommand{\Biggg}{\bBigg@{3.5}}

\makeatother

\begin{document}
\title{Unified picture of superconductivity and magnetism in CeRh$_2$As$_2$}
\author{Changhee Lee}
\affiliation{Department of Physics and MacDiarmid Institute for Advanced Materials and Nanotechnology, University of Otago, P.O. Box 56, Dunedin 9054, New Zealand}
\author{Daniel F. Agterberg}
\affiliation{Department of Physics, University of Wisconsin--Milwaukee, Milwaukee, Wisconsin 53201, USA}
\author{P. M. R. Brydon}
\affiliation{Department of Physics and MacDiarmid Institute for Advanced Materials and Nanotechnology, University of Otago, P.O. Box 56, Dunedin 9054, New Zealand}

\begin{abstract}
We study the micorscopic origin of the multiple superconducting and magnetic phases observed in
CeRh$_2$As$_2$. We exploit the existence of a van Hove singularity enforced by the nonsymmorphic symmetry to conduct a renormalization group analysis. When Fermi-surface nesting is strong, we find two closely-competing superconducting states with opposite parities, as well as an instability towards specific spin-density wave states, consistent with key features of the phase diagram of CeRh$_2$As$_2$.
\end{abstract}

\maketitle

\textit{Introduction.}---The heavy-fermion superconductor CeRh$_2$A$s_2$ has been celebrated for its large critical field $H_{\rm c2}$ which far exceeds the Pauli limit, and a field-induced first-order transition between distinct low- and high-field superconducting (SC) phases~\citep{Khim2021}. These features have been widely interpreted as indicating a transition between two nearly-degenerate SC states with opposite parity, which is enabled by the locally non-centrosymmetric crystal structure~\citep{Khim2021,Sigrist2014, Mockli2021,Cavanagh2022,Nogaki2022}. Although this explanation is in quantitative agreement with the observed temperature- and magnetic-field-dependence of the superconductivity~\cite{Landaeta2022}, the phase diagram is in fact much richer: an enigmatic so-called quadrupole density wave phase appears above the SC transition temperature \citep{Hafner2022,Pfeiffer2023Expt,Mishra2022Expt,Miyake2023,Khim2024uSR}, and an antiferromagnetic phase inside the SC phase has been identified \citep{Ogata2023,Kibune2022,Khim2024uSR}. The presence of these phases cast some doubts on the interpretation of the field-induced first-order transition of ${\rm{CeRh_2As_2}}$ as a switch between distinct SC states, with several researchers instead arguing that it indicates a metamagnetic transition, where the magnetic moments are realigned by the magnetic field 
\citep{Machida2022, Chajewski2024Expt}.

The controversy surrounding the nature of the first-order transition in CeRh$_2$As$_2$ motivates us to investigate
the microscopic origin of the superconductivity and the magnetism. The usual starting point for such a study is an accurate model of the low-energy electronic structure. This is not available for CeRh$_2$As$_2$, however, because the Fermi surface structure predicted by {\it ab initio} methods is found to be extremely sensitive to how the Ce $4f$-electrons are treated \citep{Ptok2021,Hafner2022,Cavanagh2022,Ishizuka2023,Ma2024}.
To clarify the situation, the low-energy electronic structure has been investigated by several angle-resolved photoemission (ARPES) experiments. These studies consistently report a quasi-2D Fermi surface and a high-intensity signal around the X points of the first Brillouin zone (FBZ) \citep{BChen2024Expt,Wu2024Expt,XChen2024PRX}, which have been interpreted as evidence of van Hove singularities (VHSs) near the Fermi energy.

The existence of VHSs in quasi-two-dimensional electronic bands is highly significant, as their singular contribution to the density of states allows the interplay between magnetism and superconductivity to be understood in terms of the nesting between VHSs. The physics of such systems is most elegantly exploited within the parquet renormalization group (RG) method \citep{Zheleznyak1997,Schulz1987,Maiti2013}. The competition between unconventional SC phases and antiferromagnetism in cuprates, graphene, and Kagome materials \citep{Furukawa1998,Hur2009,Nandkishore2012,Kiesel2013Kagome,Lin2021Kagome} is explained by parquet RG in terms of VHSs at time-reversal invariant momenta (TRIM), so-called type-I VHSs \citep{Yao2015}. Type-II VHSs, which are positioned off the TRIM, are also known to be able to promote odd-parity superconductivity as well as exotic density waves \citep{Yao2015,Lin2020,Meng2015,Trott2020}, and have been proposed to control the physics in superconductors such as twisted bilayer graphene and rhombohedral trilayer graphene \citep{Chichinadze2020Graphene,Lu2022Graphene,You2022Graphene}.

In this work, we present a unified picture of the superconductivity and magnetism in $\rm{CeRh_2As_2}$ based on a parquet RG analysis of instabilities driven by type-II VHSs. We commence by demonstrating that the nonsymmorphic space group symmetry of CeRh$_2$As$_2$ generically leads to the presence of type-II VHSs. We then generalize the usual parquet RG analysis to account for the intersublattice hopping and Rashba spin-orbit coupling (SOC) expected in locally noncentrosymmetric systems. The resulting phase diagram in the weak-coupling limit exhibits two nearly degenerate SC instabilities with opposite parities when the Rashba SOC or the nesting of VHSs is strong. At strong nesting we also find instabilities toward spin-density wave (SDW) states, one of which is consistent with key features of the experimentally observed antiferromagnetism in $\rm{CeRh_2As_2}$. We conclude our work by relating our result with experiments.

\label{sec2}
\textit{Symmetry-enforced Type-II VHS.}---A type-II VHSs can be generated by moving a type-I VHSs off the TRIM. Such a shift of the position of the VHSs will occur if there are terms linear in $\boldsymbol{k}$ around the TRIM, as demonstrated simply by $k^{2}_{x}-k^{2}_y\rightarrow k^{2}_x- k^{2}_y-a k_x$. A linear term, however, is not possible in centrosymmetric systems subject to symmorphic space groups, where the electronic states transforms like the usual spin-1/2 fermions under crystal symmetries \citep{Kozii2015}. Schur's lemma dictates that the inversion symmetry should be represented by the trivial $2 \times 2$ matrix at TRIMs in this case \citep{Cornwell1997,Bradley2009}. 
The trivial representation of inversion symmetry, along with time-reversal symmetry, forbids any term linear in $\boldsymbol{k}$. In contrast, terms linear in $\boldsymbol{k}$ are allowed in non-symmorphic space group as we demonstrate below.

We consider a  ${\boldsymbol{k}}\cdot{\boldsymbol{p}}$ Hamiltonian at the X point of $\rm{CeRh_2As_2}$ containing the essential ingredients for our theory:
\begin{equation}
h_{\boldsymbol{k}}=(k_{y}^{2}-k_{x}^{2})\tau_{0} s_{0}+t_{x}k_{x}\tau_{x} s_{0}+\tau_{z}(\lambda_{x}k_{y} s_{x}+\lambda_{y}k_{x} s_{y}),\label{eq:H-kp}
\end{equation}
which is identical to that presented in Ref. \citep{Suh2023} up to unitary transformation. Here, $\boldsymbol{k}$ is the momentum measured from the X point in the FBZ. $\tau_\mu$ ($s_\mu$) with $\mu=0,x,y,z$ are Pauli matrices encoding the sublattice (spin) degrees of freedom. The coefficents of the nontrivial terms, $t_{x}$, $\lambda_{x}$, and $\lambda_{y}$ can be related to spin-independent sublattice hopping and spin-orbit couplings, respectively. We choose the trivial part of the Hamiltonian to display a type-I VHS at the X point when the nontrivial terms are vanishing.

The conversion of the type-I VHS into type-II VHSs due to the linear terms can be directly seen when we examine the eigenenergies of $h_{\boldsymbol{k}}$
\begin{equation}
\epsilon_{\pm}(\boldsymbol{k})=(k_{y}^{2}-k_{x}^{2})\pm\sqrt{(t_{x}^{2}+\lambda_{y}^{2})k_{x}^{2}+\lambda_{x}^{2}k_{y}^{2}},\label{eq:X-dispersion}
\end{equation}
with $+$($-$) for the conduction (valence) band. 
The saddle points of $\epsilon_{\pm}$ are found at $(\pm k_{+}, 0)$ and $( 0,\pm k_{-})$ with
\begin{equation}
|k_{+}|=\frac{1}{2}\sqrt{t_{x}^{2}+\lambda_{y}^{2}},\quad|k_{-}|=\frac{1}{2}|\lambda_{y}|.
\end{equation}
Consequently, the type-I VHS at the X point is moved off the TRIM and converted into type-II VHSs when the terms linear in $\boldsymbol{k}$ are present in Eq. \eqref{eq:H-kp}.

We would like emphasize that there is a single available irreducible representation (irrep) for spinful fermions at the X point in the space group No. 129 (P4/nmm) \cite{Elcoro2017Bilbao}, and thus ${\boldsymbol{k}}\cdot{\boldsymbol{p}}$ Hamiltonian at the X point is uniquely given by $h_{\boldsymbol{k}}$ up to unitary transformation. Moreover, the non-trivial irreducible representation $P_X=\tau_y s_0$ of the inversion symmetry at the X point, which is a necessary consequence of the glide symmetry in the nonsymmorphic space group, is essential for enabling the terms linear in $\boldsymbol{k}$. Therefore, type-II VHSs are expected to exist in $\rm{CeRh_2As_2}$, and more generally in a large range of nonsymmorphic space groups.

\begin{figure}
\includegraphics[width=0.9\linewidth]{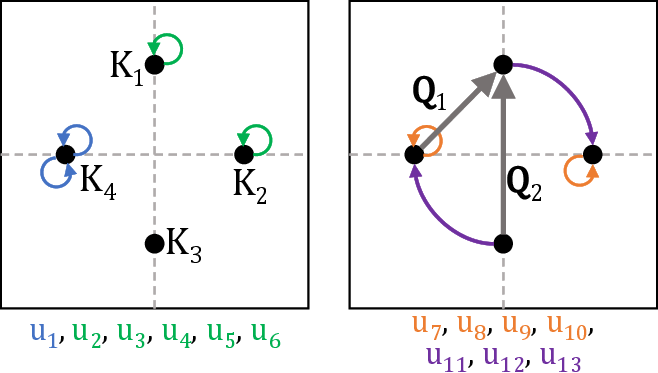}\caption{\label{fig1}Interactions between patches $\boldsymbol{K}_i$ with $i=1,\ldots,4$ centering the type-II van Hove singularities. The thick gray arrows mark the representative nesting momentum $\boldsymbol{Q}_{i}$. The particle-hole pairs involved in an interaction are depicted by the color-coded arc shaped arrows and the associated interactions are written in the same color below the figure.}
\end{figure}

\label{sec3} 
\textit{RG procedure.}---The instability of the normal phase in the presence of the type-II VHSs is investigated through the parquet RG approach \citep{Chubukov2008,Maiti2013,Lin2020}. To be consistent with ARPES results \citep{BChen2024Expt,Wu2024Expt,XChen2024PRX}, we consider the case that the Fermi level is close to the VHSs around the X and Y points in the FBZ. The divergent density of states at VHSs also permits us to focus on the interactions between electronic states in the patches around the type-II VHSs that are depicted by black dots at $K_i$ with $i=1,2,3,4$ in Fig.~\ref{fig1}. Interactions scatter particles between the different patches, as shown by the coloured arrows in this figure. The breaking of spin-rotation symmetry by the SOC results in thirteen symmetry-allowed interactions $u_i$ ($i=1,\dots,13$) between the patches \citep{SM}; this reduces to only six interactions when SOC is absent \citep{Yao2015}.

\begin{figure}
\begin{raggedright}
\includegraphics[width=1\columnwidth]{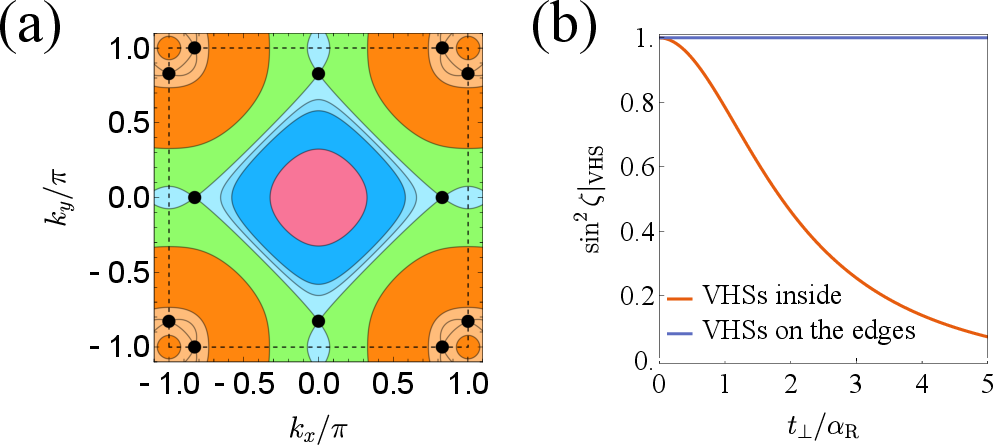}
\par
\end{raggedright}
\caption{
\label{fig2} 
(a) The energy contour map of the conduction band derived from $H(\boldsymbol{k})$ in Eq. \eqref{eq:H-LNCS} in $k_{x}-k_{y}$ plane. Black dots denote the type-II van Hove singularities. For this illustration, $\varepsilon_{x,\boldsymbol{k}}=2t_{\perp}\cos \frac{k_{x}}{2} \cos \frac{k_{y}}{2}$, ${\boldsymbol{\alpha}}_{\boldsymbol{k}}=2\alpha_{{\rm R}}(-\sin k_{y},\sin k_{x})$, and $\varepsilon_{0,{\boldsymbol{k}}}=-2t(\cos k_{x}+\cos k_{y})-\mu$ are used with $\alpha_{R}=0.6t$, $t_\perp=0.1t$.
(b) $\sin^{2}{\zeta}$ evaluated at type-II van Hove singularities as an evolution of increasing $t_{\perp}/{{\alpha}_{\rm{R}}}$ at $\alpha_{\rm{R}}=0.6t$.
}
\end{figure}

The bare interaction in our system is assumed to be a Hubbard repulsion of strength $U$ on the Ce sites. We derive the bare values of the $u_i$ by projecting the Hubbard interaction onto the low-energy subspace defined by the electronic states at the VHSs, which are obtained by using a minimal two-dimensional Hamiltonian consistent with the symmetry of $\rm{CeRh_2As_2}$ 
\citep{Nogaki2022,Cavanagh2022,Suh2023,Yufei2023}:
\begin{align}
H(\boldsymbol{k}) =&\varepsilon_{0,\boldsymbol{k}}\tau_{0}s_{0}+\varepsilon_{x,\boldsymbol{k}}\tau_{x}s_0+\boldsymbol{\alpha}_{\boldsymbol{k}}\cdot(s_{x},s_{y})\tau_{z}.\label{eq:H-LNCS}
\end{align}
Here, $\varepsilon_{0,\boldsymbol{k}}$, $\varepsilon_{x,\boldsymbol{k}}$, and $\boldsymbol{\alpha}_{\boldsymbol{k}}=(\alpha_{x,\boldsymbol{k}},\alpha_{y,\boldsymbol{k}})$ describe the intrasublattice hopping, the intersublattice hopping, and the intrasublattice Rashba SOC, respectively. The $s_\mu$ have the same meaning as in Eq.~\eqref{eq:H-kp}, but here we specify $\tau_{\mu}$ to refer to the sublattices of the $\rm{Ce}$ atoms.
Figure \ref{fig2}(a) displays the energy contour map of the conduction band for a typical tight-binding realization of $H(\boldsymbol{k})$ exhibiting type-II VHSs around the X and Y points. Given the ARPES result \citep{BChen2024Expt,Wu2024Expt,XChen2024PRX}, the additional symmetry-enforced type-II VHSs around the M point are assumed to be far away from the Fermi level, and thus we ignore them in the following.

The bare values of coupling constants $u_i$ obtained by this projection typically depend on the quantity
\begin{equation}
\sin^2{\zeta}=\frac{\boldsymbol{\alpha}_{\boldsymbol{k}}^2}{\varepsilon_{x,\boldsymbol{k}}^2+\boldsymbol{\alpha}_{\boldsymbol{k}}^2}
\end{equation}
evaluated at VHSs. The angle $\zeta$ parameterizes the relative importance of the Rashba SOC and the intersublattice hopping.
The value of $\sin^2{\zeta}$ at the VHS inside the FBZ is dependent on the microscopic parameters, but it is fixed to 1 if the VHSs are located on the edges of the FBZ, because the intersublattice hopping is vanishing on the edges due to the non-symmorphic symmetries. This feature is demonstrated in Fig. \ref{fig2}(b). We note that our result does not depend on whether we take the conduction band or the valence band for analysis because of the chiral symmetry $\tau_y$ of the non-trivial part of $H(\boldsymbol{k})$ which also leaves on-site interactions unchanged. 

\begin{figure}\includegraphics[width=1\linewidth]{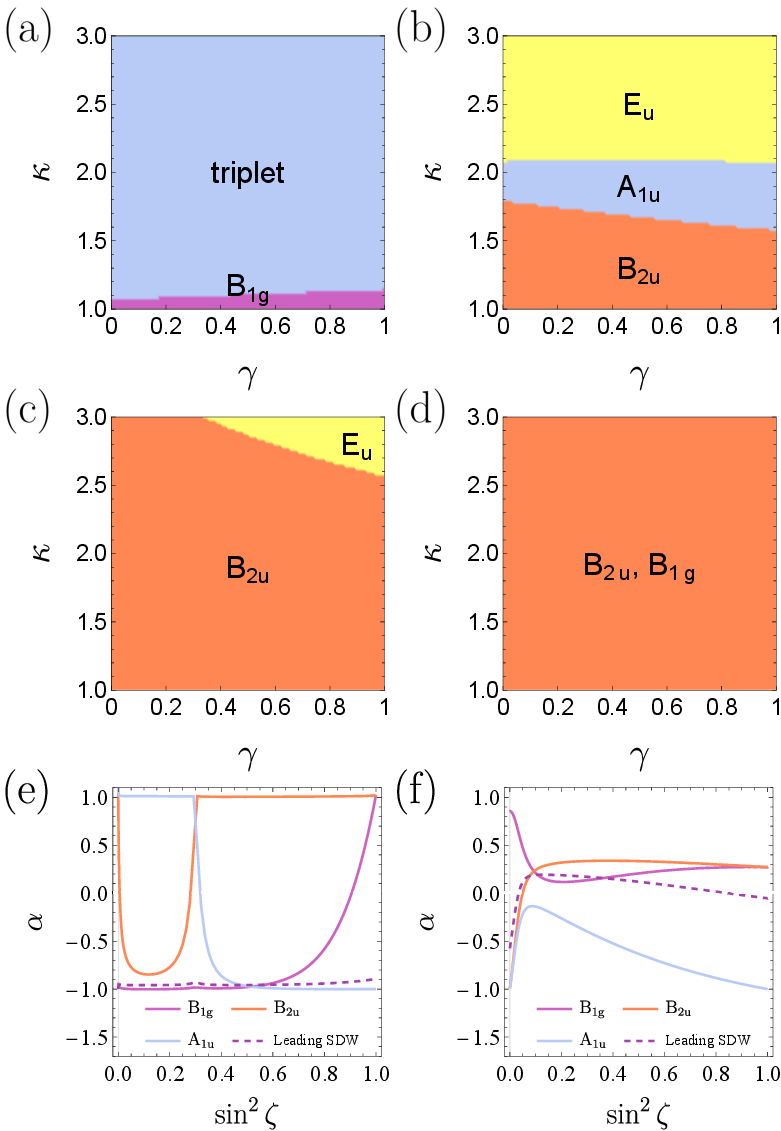}\caption{\label{fig3} (a)-(d) Weak-coupling phase diagrams in $\gamma-\kappa$ plane obtained with $\sin^{2}\zeta=0,\;0.2,\;0.5,\;1$, 
respectively, at $U=0.02$. Each phase is labeled by the symmetry of the gap function while the {\it{triplet}} phase is a phase with all spin-triplet pairings are degenerate. (e)-(f) The exponents $\alpha$ of susceptibilities of the leading and the subleading SC channels (solid line) and the leading spin-density wave channels (dashed line) at $(\kappa,\gamma)=(2,0.8)$ and $(\kappa,\gamma)=(1.001,0.8)$ for (e) and (f), respectively. The exponents are evaluated at $y=0.95y_{c}$.}\end{figure}

The four VHSs are connected by the nesting vectors  ${\boldsymbol{Q}}_{1}\equiv{\boldsymbol{K}}_{1}-{\boldsymbol{K}}_{4}$ and ${\boldsymbol{Q}}_{2}={\boldsymbol{K}}_{1}-{\boldsymbol{K}}_{3}$, as depicted by the grey arrows in Fig.~\ref{fig1}. For brevity, we also denote the nesting momentum obtained by rotating ${\boldsymbol{Q}}_{i}$ by the same symbol unless otherwise noted. Taking the approximate energy dispersion $\varepsilon_{\boldsymbol{k}}=\frac{k_{x}^2}{2m_x}-\frac{k_{y}^2}{2m_y}$ and $\varepsilon_{(k_y,k_x)}$ for the low-energy states at ${\boldsymbol{K}}_{1,3}$ and ${\boldsymbol{K}}_{2,4}$, respectively, we follow Refs. \citep{Lin2020,Yao2015,Trott2020} and measure the extent of the nesting associated with ${\boldsymbol{Q}}_{1}$ by $\kappa=m_{y}/m_{x}$. Perfect nesting in the particle-hole channel occurs at $\kappa=1$. The nesting associated with ${\boldsymbol{Q}}_{2}$ is parameterized by $0\le\gamma\le1$: The larger the value of $\gamma$, the more favorable are the particle-particle channels of the pair density waves with total momentum ${\boldsymbol{Q}}_2$ and the particle-hole channels of density waves with propagation vector ${\boldsymbol{Q}}_{2}$.

Following the standard parquet RG procedure \citep{Nandkishore2012,Maiti2013,Lin2020}, we derive the RG equations of the thirteen coupling constants by computing the one-loop diagrams and keeping only the $\mathcal{O}(\ln\frac{\Lambda}{E})$ and $\mathcal{O}(\ln^{2}\frac{\Lambda}{E})$ corrections, where the ultraviolet cutoff $\Lambda$ and the probing energy scale $E$ can be replaced by the band separation at the VHSs and the temperature, respectively. These assumptions let us use $y\equiv\ln^{2}\frac{\Lambda}{E}$ as the running parameter of the RG flow. Solving the RG equations, we find that the  coupling constants diverge at a critical scale denoted by $y_{c}$. We identify the instability of the normal phase by considering singularities in the susceptibilities $\chi$ for all possible particle-particle channels and particle-hole channels in accordance with our patch approximation.

Near $y=y_c$, some of the susceptibilities are found to exhibit singular behaviour $\sim (y_c -y)^{-\alpha}$, where a positive value of the exponent $\alpha$ indicates an instability, and the channel with the largest positive $\alpha$ is regarded as the leading instability \citep{Lin2020,Maiti2013}. The patch approximation can lead to spurious degeneracies of different orders in the particle-hole channels, which should be lifted by considering the full Fermi surface. We refer the reader to the Supplemental material for further details on our RG method \cite{SM}.

\textit{Superconducting instability}-- Figure \ref{fig3} shows the weak-coupling phase diagrams in the $\gamma-\kappa$ plane for several values of $\zeta$ at $U=0.02$. We find leading instabilities only among particle-particle channels with zero total momentum, which are referred to as SC channels; we label the leading SC phase by the irreps of the point group $D_{{\rm 4h}}$.

Let us first look into the phase diagram without spin-orbit coupling displayed in Fig. \ref{fig3}(a). Consistent with Ref. \citep{Yao2015}, two SC phases are found, depending mainly on the value of $\kappa$.
When $\kappa$ is low, the strong particle-hole nesting with momentum  $\boldsymbol{Q}_{1}$ promotes $d$-wave Cooper pairing, denoted by $B_{{\rm 1g}}$. Upon increasing $\kappa$, spin-triplet pairing takes over from the $d$-wave pairing as the nesting is reduced. All the spin-triplet pairings are degenerate as spin-orbit coupling is absent.

For a weak SOC with $\sin^{2}\zeta=0.2$ [Fig. \ref{fig3}(b)], the $B_{{\rm 1g}}$ phase is quickly suppressed and becomes subdominant, while the aforementioned spin-triplet phase is split into three SC phases. A $B_{{\rm 2u}}$ phase appears at low $\kappa$ taking the place of the $B_{\rm{1g}}$ SC phase. We can understand this result as follows: Although the Rashba SOC is expected to favour an $A_{\rm{2u}}$ state with ${\bf d}$-vector aligned parallel to the spin texture \citep{Cavanagh2022}, this is modulated by a $B_{{\rm 1g}}$ form factor due to strong $\boldsymbol{Q}_{1}$ particle-hole nesting, giving a $B_{{\rm 2u}}=B_{{\rm 1g}}\times A_{{\rm 2u}}$ SC phase. When $\kappa$ is very large so that the $\boldsymbol{Q}_{1}$ nesting is weak, patches on the $x$-axis and $y$-axis become essentially independent of each other. As a result, the $E_{\rm u}$ SC phase with two degenerate gap functions becomes the dominant channel, and will realize either a nematic or chiral SC phase. An $A_{\rm{1u}}$ SC phase appears at intermediate $\kappa$. 

Further increasing the Rashba SOC, the $B_{{\rm 2u}}$ SC phase appears at larger $\kappa$ as shown in Fig. \ref{fig3}(c). This trend continues to the limit of $\sin^{2}\zeta=1$ [Fig. \ref{fig3}(d)], where the $B_{2u}$ SC phase is found for all $\kappa$ and is degenerate with the $B_{{\rm 1g}}$ SC channel. This degeneracy arises from the vanishing effect of intersublattice hopping at $\sin^{2}\zeta=1$ which makes both layers independently subject to the non-centrosymmetric point group $C_{{\rm 4v}}$, of which the gap functions of $B_{{\rm 2u}}$ SC state and $B_{{\rm 1g}}$ SC state belong to the same irrep \citep{Nogaki2022}. The other odd-parity pairings are completely suppressed at $\sin^{2}\zeta=1$ because they involve only interband pairing in this limit. Although the $B_{{\rm 2u}}$ and $B_{{\rm 1g}}$ are only degenerate at $\sin^{2}\zeta=1$, this nevertheless leads them to be closely competing at high relative SOC strength or strong $\boldsymbol{Q}_{1}$ nesting, as shown in Figs. \ref{fig3}(e) and \ref{fig3}(f). The close competition of even- and odd-parity SC states over a wide parameter regime is consistent with a field-induced parity-transition in CeRh$_2$As$_2$. The specific combination of $B_{{\rm 2u}}$ and $B_{{\rm 1g}}$ leading instabilities has been previously identified in a fluctuation-exchange calculation for a tight-binding model \cite{Nogaki2022}.

\label{sec4}
\textit{Spin Density Wave.}---Although the leading instabilities of our model are always in the SC channels in the weak coupling limit with $\kappa>1$, we find subleading instabilities in one set of degenerate particle-hole channels. These channels correspond to time-reversal symmetry-breaking density waves with a propagation vector ${\boldsymbol{Q}}_{1}=\pm({\boldsymbol{K}}_1-{\boldsymbol{K}}_4)$ or $\bar{\boldsymbol{Q}}_{1}=\pm({\boldsymbol{K}}_1-{\boldsymbol{K}}_2)$. Hereafter, we refer to them as the leading spin-density waves (SDWs) for short. All other types of density waves, including  those with a propagation vector ${\boldsymbol{Q}}_2$ or 
Pomeranchuk-like instabilities such as the odd-parity magnetic multipolar order~\citep{Kibune2022}, never appear with positive exponent $\alpha$. 

We plot the exponent of the leading SDW channels in Figs. \ref{fig3}(e)-(f) as the dashed line. 
When nesting is weak, as shown in Fig. \ref{fig3}(e), the exponent of SDW channels are $\alpha \approx -1$, and there is no instability towards the SDW phases.
At stronger nesting, however, the leading SDWs become comparable with the SC channels for a range of $\sin^{2}\zeta$, as shown in Fig. \ref{fig3}(f), although they never become the leading instability. The range of  $\sin^{2}\zeta$ for which the SDW phases show an instability increases with $U$, as presented in the Supplemental Material~\cite{SM}.
Consequently, we conclude that a subdominant SDW instability is possible when the nesting is strong.

\begin{figure}
\includegraphics[width=0.9\columnwidth]{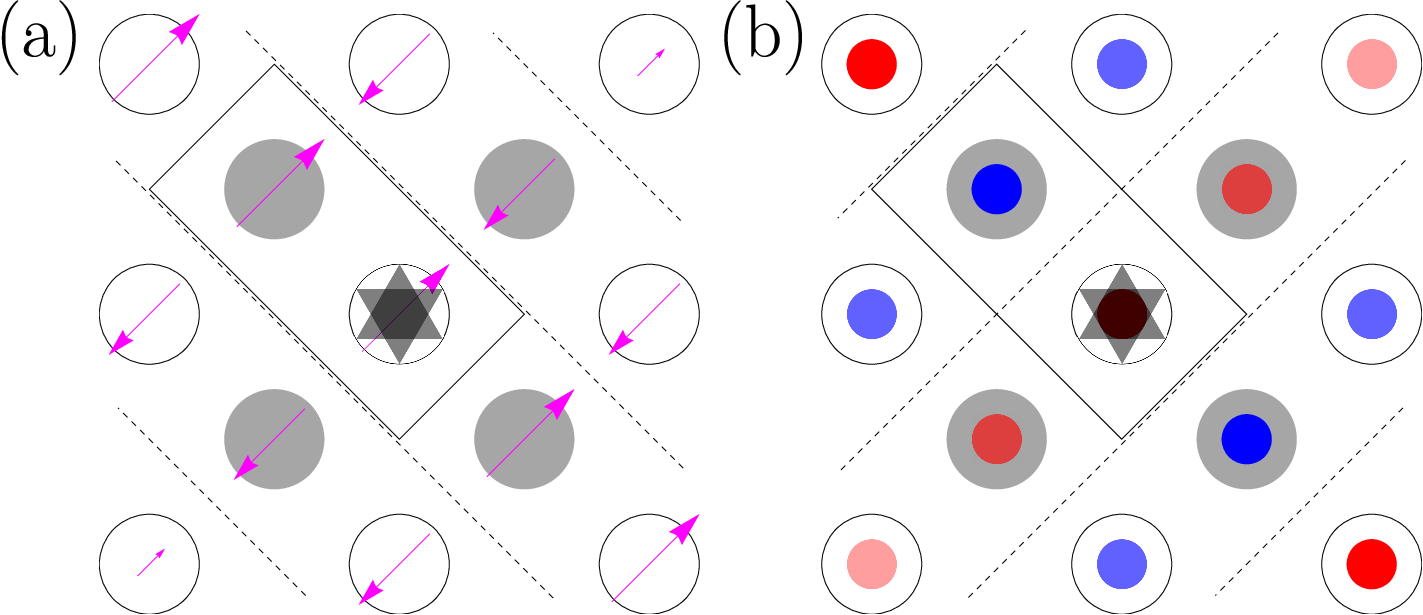}\caption{\label{fig4}
Top view of the local magnetic moments of (a) $M_{xy,\parallel}$ and (b) $M_{z,\perp}$ with propagation vector ${\boldsymbol{Q}}_1=0.8(\pi,\pi)$ given in Eq. \eqref{eq:B3gSDW} and Eq. \eqref{eq:AuSDW}, respectively. Filled (empty) disks correspond to the sublattice A(B) of the Ce lattice in $\rm{CeRh_2As_2}$. Arrows and red (blue) dots represents the in-plane and positive (negative) out-of-plane magnetic moments, respectively. A primitive unit cell is depicted by a black box. Four-fold rotations are conducted along the axis passing through the black star. 
}  
\end{figure}

We find that the leading SDW is consistent with two distinct arrangements of local magnetic moments on the Ce sites:
\begin{align}
    M_{xy,\parallel}&:{\boldsymbol{m}}_{A(B)}\propto    (\hat{\boldsymbol{x}}+\hat{\boldsymbol{y}})
    \cos ({\boldsymbol{Q}}_{1} \cdot {\boldsymbol{R}}_{\rm A(B)}+\varphi),\label{eq:B3gSDW}\\
    M_{z,\perp}&:{\boldsymbol{m}}_{A(B)}\propto \pm\hat{\boldsymbol{z}}
    \cos ({\boldsymbol{Q}}_{1} \cdot  {\boldsymbol{R}}_{\rm A(B)}+\varphi),\label{eq:AuSDW}
\end{align}
with $+$($-$) in Eq. \eqref{eq:AuSDW} for the sublattice A (B) of Ce atoms and $0\le\varphi<\pi$. Figure \ref{fig4}(a) and \ref{fig4}(b) illustrate the local magnetic moments of $M_{xy,\parallel}$ and $M_{z,\perp}$ with $\varphi=0$ and the propagation vector ${\boldsymbol{Q}}_{1}=0.8(\pi,\pi)$, respectively.
$M_{xy,\parallel}$ describes a longitudinal wave of in-plane magnetic moments, where the local moments in the stripes perpendicular to the propatagion vector are aligned in the same direction, either parallel or antiparallel to the propatagion vector. On the other hand, $M_{z,\perp}$ is a transverse SDW of out-of-plane magnetic moments whose direction alternates in a stripe as shown in Fig. \ref{fig4}(b). Each of these states is degenerate with an SDW with a propagation vector $\bar{\boldsymbol{Q}}_{1}$, obtained by rotating each of them $90^\circ$ about the axis defined by the star in Fig. \ref{fig4}. This raises the possibility of a double-Q SDW phase, similar to that found in the iron-based superconductors \citep{Allred2016,JWang2017}: a double-Q $M_{xy,\parallel}$ would realize a hedgehog-like spin-vortex texture, whereas the double-Q $M_{z,\perp}$ corresponds to a spin-charge order \citep{Meier2018,Halloran2017}. Representative plots of these orders are provided in the SM \citep{SM}. 

Remarkably, $M_{xy,\parallel}$ is consistent with what is known of the antiferromagnetic state in CeRh$_2$As$_2$. In addition to featuring in-plane local magnetic moments \citep{Kitagawa2022,TChen2024Expt}, it generates finite magnetic fields at the As sites in the As-Rh-As block but not in the Rh-As-Rh block \citep{Kibune2022,Ogata2023}. To be consistent with neutron scattering \cite{TChen2024Expt}, ${\boldsymbol{Q}}_1$ must be close to $(\pi,\pi)$, i.e. the type-II VHSs are near to the X and Y points, which is in agreement with the location of the VHSs suggested by ARPES.
On the other hand, the SDW $M_{z,\perp}$ of out-of-plane local moments produces magnetic fields at all As site, in disagreement with experimental observations \citep{Kibune2022,Ogata2023,Kitagawa2022,TChen2024Expt}.

\label{sec5}
\textit{Conclusion.}---We used the parquet RG approach, which treats  superconducting and density-wave instabilities on an equal footing, to investigate the normal phase instabilities of $\rm{CeRh_{2}As_{2}}$ driven by the interactions between the symmetry-enforced type-II VHSs near the X and Y points of the FBZ. We find two closely-competing SC instabilities of opposite parities, but also  subdominant SDW instabilities when the nesting is strong. Of the two SDWs we find, the in-plane and longitudinal SDW of local magnetic moments captures characteristic features of the antiferromagnetic state observed in recent experiments \citep{Kibune2022,Ogata2023,Kitagawa2022,TChen2024Expt}.

Our study presents a unified picture of the physics of CeRh$_2$As$_2$, supporting the parity-transition scenario for the multiple SC phases, and implying an intimate link between the SC and antiferromagnetism. Our results may also be relevant to the quadrupole density wave, which recent $\mu$SR experiments suggests has a magnetic origin~\cite{Khim2024uSR}, 
but greater clarity on the microscopic nature of this phase is required to make a definitive statement.

{\it Acknowledgements.---} CL and PMRB were supported by the
Marsden Fund Council from Government funding, managed by Royal Society Te Ap\={a}rangi, Contract No. UOO2222. We acknowledge useful discussions with Tatsuya Shishiduo, Mike Weinert, and Yue Yu.

\bibliography{ref}
\end{document}


\global\long\def\thefigure{S\arabic{figure}}%

\title{Supplemental Material:\\Unified picture of the superconductivity and magnetism in CeRh$_2$As$_2$}
\author{Changhee Lee}
\affiliation{Department of Physics and MacDiarmid Institute for Advanced Materials
and Nanotechnology, University of Otago, P.O. Box 56, Dunedin 9054,
New Zealand}
\author{Daniel F. Agterberg}
\affiliation{Department of Physics, University of Wisconsin--Milwaukee, Milwaukee, Wisconsin 53201, USA}
\author{P. M. R. Brydon}
\affiliation{Department of Physics and MacDiarmid Institute for Advanced Materials
and Nanotechnology, University of Otago, P.O. Box 56, Dunedin 9054,
New Zealand}

\maketitle
\tableofcontents{}

\section{Pseudospin basis}

The bilayer of locally non-centrosymmetric (LNCS) systems we study
in the main text is described by

\begin{align}
H_{{\rm TB}}^{({\rm LNCS})}(\boldsymbol{k}) & =\varepsilon_{0,\boldsymbol{k}}I_{4}+\varepsilon_{x,\boldsymbol{k}}\tau_{x}+\boldsymbol{\alpha}_{\boldsymbol{k}}\cdot(s_{x},s_{y})\tau_{z}\label{eq:H-LNCS}
\end{align}
where $\tau_{\mu}$ and $s_{\mu}$ with $\mu=0,x,y,z$ are the Pauli
matrices of the sublattice and the spin degrees of freedom, respectively,
and the inversion and the time reversal symmetries are represented
by $\tau_{x}$ and $is_{y}{\cal K}$, respectively, with the complex
conjugation ${\cal K}$. In the square lattice, the intralayer Rashba
SOC and the intralayer hopping are given by $\boldsymbol{\alpha}_{\boldsymbol{k}}=(\alpha_{\boldsymbol{k},x},\alpha_{\boldsymbol{k},y})=2\alpha_{{\rm R}}(-\sin k_{y},\sin k_{x}),$
$\varepsilon_{0,\boldsymbol{k}}=2t(\cos k_{x}+\cos k_{y})+4t_{nn}\cos k_{x}\cos k_{y}-\mu$,
respectively. Regarding the interlayer hopping $\varepsilon_{x,\boldsymbol{k}}$,
it can be $2t_{\perp}$ or $2t_{\perp}\cos\frac{k_{x}}{2}\cos\frac{k_{y}}{2}$.
The formal case corresponds to when the lattice is subject to a symmorphic
space group while the latter describes when a non-symmorphic group
is in action. For the completeness, we also introduce here
\begin{equation}
\sin^{2}\zeta=\frac{\boldsymbol{\alpha}_{\boldsymbol{k}}^{2}}{\varepsilon_{x,\boldsymbol{k}}^{2}+\boldsymbol{\alpha}_{\boldsymbol{k}}^{2}}\bigg|_{{\rm VHS}},\label{eq:sinzeta}
\end{equation}
which is the same as defined in the main text.

The eigenvectors of $H_{{\rm TB}}^{({\rm LNCS})}(\boldsymbol{k})$
are given in a matrix form \citep{Cavanagh2022}:
\begin{align}
U(\boldsymbol{k})= & (U_{{\rm c}}(\boldsymbol{k}),U_{v}(\boldsymbol{k}))=\left(\begin{matrix}\cos\frac{\chi}{2}\cos\frac{\omega}{2} & -e^{-i\phi}\sin\frac{\chi}{2}\sin\frac{\omega}{2} & -i\cos\frac{\chi}{2}\sin\frac{\omega}{2} & ie^{-i\phi}\sin\frac{\chi}{2}\cos\frac{\omega}{2}\\
-e^{i\phi}\sin\frac{\chi}{2}\cos\frac{\omega}{2} & -\cos\frac{\chi}{2}\sin\frac{\omega}{2} & ie^{i\phi}\sin\frac{\chi}{2}\sin\frac{\omega}{2} & i\cos\frac{\chi}{2}\cos\frac{\omega}{2}\\
-\cos\frac{\chi}{2}\sin\frac{\omega}{2} & e^{-i\phi}\sin\frac{\chi}{2}\cos\frac{\omega}{2} & i\cos\frac{\chi}{2}\cos\frac{\omega}{2} & ie^{-i\phi}\sin\frac{\chi}{2}\sin\frac{\omega}{2}\\
e^{i\phi}\sin\frac{\chi}{2}\sin\frac{\omega}{2} & -\cos\frac{\chi}{2}\cos\frac{\omega}{2} & ie^{i\phi}\sin\frac{\chi}{2}\cos\frac{\omega}{2} & i\cos\frac{\chi}{2}\sin\frac{\omega}{2}
\end{matrix}\right)_{\omega=\frac{\pi}{2}}\label{eq:pseudospin}
\end{align}
where the first (last) two columns denoted by $U_{{\rm c}}(\boldsymbol{k})$
($U_{{\rm v}}(\boldsymbol{k})$) correspond to the eigenstates of
the conduction (valence) band, whose dispersion is given by $E_{+}(\boldsymbol{k})=\varepsilon_{0,\boldsymbol{k}}+\sqrt{\varepsilon_{x,\boldsymbol{k}}^{2}+|\boldsymbol{\alpha}_{\boldsymbol{k}}|^{2}}$
($E_{-}(\boldsymbol{k})=\varepsilon_{0,\boldsymbol{k}}-\sqrt{\varepsilon_{x,\boldsymbol{k}}^{2}+|\boldsymbol{\alpha}_{\boldsymbol{k}}|^{2}}$).
Here, $\chi=\arctan\frac{\varepsilon_{x,\boldsymbol{k}}}{|\alpha_{\boldsymbol{k}}|}$
and $\phi=\arctan\frac{\alpha_{\boldsymbol{k},y}}{\alpha_{\boldsymbol{k},x}}$.
The two eigenstates of each band form the so-called pseudospin basis
by which the symmetry operators are represented in the same way as
the ${\rm SU(2)}$ spin rotation matrices. For instance, the representation
of the mirror reflection symmetry $M_{100}$ against $yz$-plane is
given in this pseudospin basis as 
\begin{align}
U_{{\rm v}}^{\dagger}(M_{100}^{-1}\boldsymbol{k})D(M_{100})U_{{\rm v}}(\boldsymbol{k})=\exp(-i\frac{\pi}{2}\sigma_{x})
\end{align}
with the pseudospin Pauli matrix $\sigma_{x}$ along the $x$-axis.

For the van Hove singularities on the edges of the FBZ, the interlayer hopping vanishes at these points due to the non-symmorphic symmetry. As a result, the symmetry of the system is effectively reduced into a direct sum of two symmorphic groups with the $C_{\rm{4v}}$ point group for each sublattice of Ce site. Therefore, this case can be guessed by the result obtained from the patch model with patches inside the FBZ and $\varepsilon_{x,\boldsymbol{k}}=0$.

\section{Interactions and Renormalization group equations}
\label{sec:Interactions}

\subsection{Interactions when spin-rotation symmetry is broken}

\begin{figure}
\includegraphics[width=0.98\textwidth]{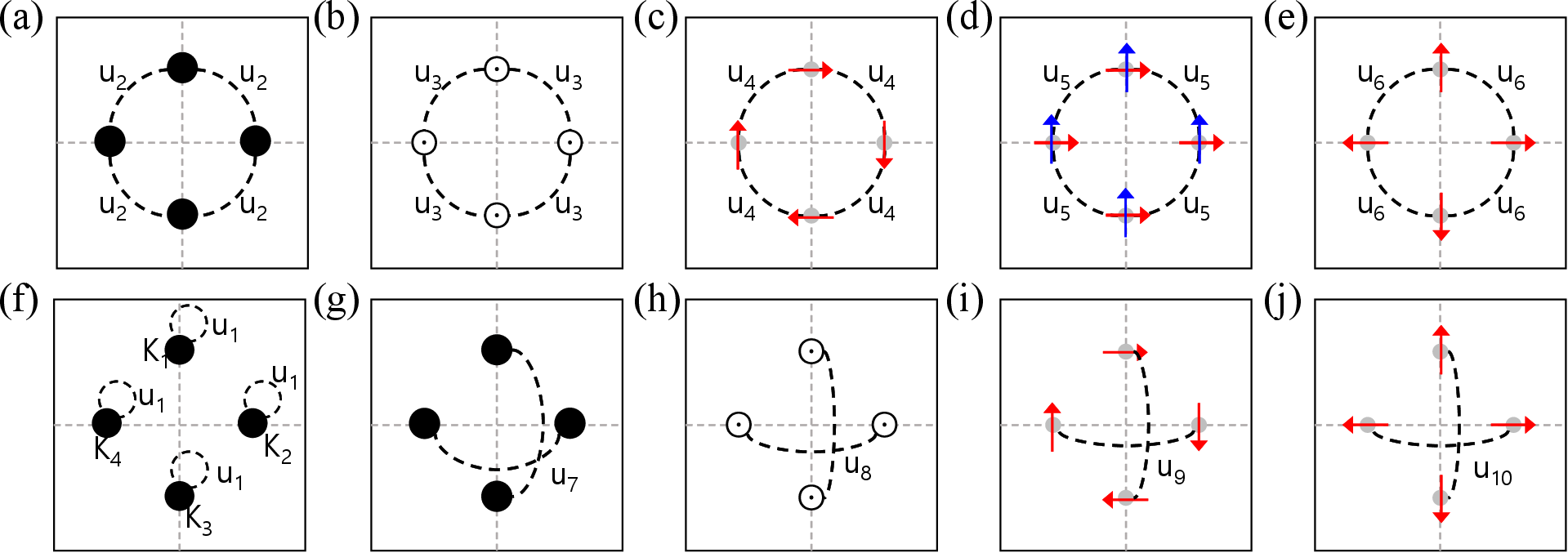}\caption{\label{fig:S1}Description of the interactions $u_{1}$-$u_{10}$
in the square lattice in the particle-hole channel by means of pseudospin
moment. Patches are marked by small gray disks. Black circles denote
the density-density type interactions. The out-of-plane pseudospin
moment and in-plane pseudospin moments are marked by $\odot$ and
arrows. Dashed lines links the patches involved in the scattering
due to the associated interaction.}
\end{figure}

We provide the details of the interactions allowed by the crystallographic
space group in the presence of the spin-orbit coupling. As mentioned
in the main text, broken spin-rotation symmetry allows four types
of scatterings containing 13 interactions between the patches around
type-II van Hove singularities (VHSs) regardless of whether the VHSs
are located on the edge of or inside the first Brillouin zone (BZ).
The pseudospin representation of these 13 interactions are given by

\begin{align}
u_{1}: & \sum_{i=1,\cdots,4}(\psi_{K_{i}}^{\dagger}\sigma_{0}\psi_{K_{i}})^{2}\label{eq:u1}\\
u_{2}: & \{\psi_{K_{1}}^{\dagger}\sigma_{0}\psi_{K_{1}}+\psi_{K_{3}}^{\dagger}\sigma_{0}\psi_{K_{3}}\}\{\psi_{K_{2}}^{\dagger}\sigma_{0}\psi_{K_{2}}+\psi_{K_{4}}^{\dagger}\sigma_{0}\psi_{K_{4}}\}\\
u_{3}: & \{\psi_{K_{1}}^{\dagger}\sigma_{z}\psi_{K_{1}}+\psi_{K_{3}}^{\dagger}\sigma_{z}\psi_{K_{3}}\}\{\psi_{K_{2}}^{\dagger}\sigma_{z}\psi_{K_{2}}+\psi_{K_{4}}^{\dagger}\sigma_{z}\psi_{K_{4}}\}\\
u_{4}: & \{\psi_{K_{1}}^{\dagger}\check{\sigma}_{x}\psi_{K_{1}}+\psi_{K_{3}}^{\dagger}\check{\sigma}_{x}\psi_{K_{3}}\}\{\psi_{K_{2}}^{\dagger}\check{\sigma}_{x}\psi_{K_{2}}+\psi_{K_{4}}^{\dagger}\check{\sigma}_{x}\psi_{K_{4}}\}\\
u_{5}: & \{\psi_{K_{1}}^{\dagger}\check{\sigma}_{x}\psi_{K_{1}}-\psi_{K_{3}}^{\dagger}\check{\sigma}_{x}\psi_{K_{3}}\}\{\psi_{K_{2}}^{\dagger}\check{\sigma}_{y}\psi_{K_{2}}-\psi_{K_{4}}^{\dagger}\check{\sigma}_{y}\psi_{K_{4}}\}\nonumber \\
 & +\{\psi_{K_{1}}^{\dagger}\check{\sigma}_{y}\psi_{K_{1}}-\psi_{K_{3}}^{\dagger}\check{\sigma}_{y}\psi_{K_{3}}\}\{-\psi_{K_{2}}^{\dagger}\check{\sigma}_{x}\psi_{K_{2}}+\psi_{K_{4}}^{\dagger}\check{\sigma}_{x}\psi_{K_{4}}\}\nonumber \\
u_{6}: & \{\psi_{K_{1}}^{\dagger}\check{\sigma}_{y}\psi_{K_{1}}+\psi_{K_{3}}^{\dagger}\check{\sigma}_{y}\psi_{K_{3}}\}\{\psi_{K_{2}}^{\dagger}\check{\sigma}_{x}\psi_{K_{2}}+\psi_{K_{4}}^{\dagger}\check{\sigma}_{x}\psi_{K_{4}}\}\\
u_{7}: & (\psi_{K_{1}}^{\dagger}\check{\sigma}_{0}\psi_{K_{1}})(\psi_{\bar{K}_{1}}^{\dagger}\check{\sigma}_{0}\psi_{\bar{K}_{1}})+(\psi_{K_{2}}^{\dagger}\check{\sigma}_{0}\psi_{K_{2}})(\psi_{\bar{K}_{2}}^{\dagger}\check{\sigma}_{0}\psi_{\bar{K}_{2}})\\
u_{8}: & (\psi_{K_{1}}^{\dagger}\check{\sigma}_{z}\psi_{K_{1}})(\psi_{\bar{K}_{1}}^{\dagger}\check{\sigma}_{z}\psi_{\bar{K}_{1}})+(\psi_{K_{2}}^{\dagger}\check{\sigma}_{z}\psi_{K_{2}})(\psi_{\bar{K}_{2}}^{\dagger}\check{\sigma}_{z}\psi_{\bar{K}_{2}})\\
u_{9}: & -(\psi_{K_{1}}^{\dagger}\check{\sigma}_{x}\psi_{K_{1}})(\psi_{\bar{K}_{1}}^{\dagger}\check{\sigma}_{x}\psi_{\bar{K}_{1}})-(\psi_{K_{2}}^{\dagger}\check{\sigma}_{x}\psi_{K_{2}})(\psi_{\bar{K}_{2}}^{\dagger}\check{\sigma}_{x}\psi_{\bar{K}_{2}})\\
u_{10}: & -(\psi_{K_{1}}^{\dagger}\check{\sigma}_{y}\psi_{K_{1}})(\psi_{\bar{K}_{1}}^{\dagger}\check{\sigma}_{y}\psi_{\bar{K}_{1}})-(\psi_{K_{2}}^{\dagger}\check{\sigma}_{y}\psi_{K_{2}})(\psi_{\bar{K}_{2}}^{\dagger}\check{\sigma}_{y}\psi_{\bar{K}_{2}})
\end{align}
where $\bar{\boldsymbol{K}}_{i}=-\boldsymbol{K}_{i}=\boldsymbol{K}_{i+2}$
for $i=1,2,3,4$ is understood. The summation over the momentum $\boldsymbol{k}$
which is small so that $\boldsymbol{K}_{i}+\boldsymbol{k}$ resides
inside the patch centering $\boldsymbol{K}_{i}$ is omitted for the
conciseness. Here, we introduced $\check{\sigma}_{\mu}$ a pseudospin
momentum viewed in the local frame of pseudospin space for each patch.
It is defined as $\psi_{K_{i}}^{\dagger}\check{\sigma}_{\mu}\psi_{K_{i}}=\psi_{K_{i}}^{\dagger}e^{-i\frac{\phi_{i}}{2}\sigma_{z}}\sigma_{\mu}e^{i\frac{\phi_{i}}{2}\sigma_{z}}\psi_{K_{i}}$
with $\phi_{i}=\frac{\pi}{2}(i-1)$ for each patch $K_{i}$, which
affects only the in-plane spin momentum. To be specific,
\begin{align}
\psi_{K_{1}}^{\dagger}(\check{\sigma}_{x},\check{\sigma}_{y})\psi_{K_{1}}= & \psi_{K_{1}}^{\dagger}(\sigma_{x},\sigma_{y})\psi_{K_{1}},\\
\psi_{K_{2}}^{\dagger}(\check{\sigma}_{x},\check{\sigma}_{y})\psi_{K_{2}}= & \psi_{K_{2}}^{\dagger}(-\sigma_{y},\sigma_{x})\psi_{K_{2}},\\
\psi_{K_{3}}^{\dagger}(\check{\sigma}_{x},\check{\sigma}_{y})\psi_{K_{3}}= & \psi_{K_{3}}^{\dagger}(-\sigma_{x},-\sigma_{x})\psi_{K_{3}},\\
\psi_{K_{4}}^{\dagger}(\check{\sigma}_{x},\check{\sigma}_{y})\psi_{K_{4}}= & \psi_{K_{4}}^{\dagger}(\sigma_{y},-\sigma_{x})\psi_{K_{4}}.
\end{align}

Figure \ref{fig:S1} shows the pictorial description of representative
terms contained in the interactions $u_{1}$-$u_{10}$ by using the
particle-hole representation of them. A directed curved arrow from
a patch $K_{i}$ to a patch $K_{j}$ represents a bilinear $\psi_{K_{j}}^{\dagger}\psi_{K_{i}}$.
The in-plane pseudospin momentum involved in the interaction are marked
by red in-plane arrows. As an example, Fig. \ref{fig:S1}(c) depicts
a term $(\psi_{K_{1}}^{\dagger}\sigma_{x}\psi_{K_{1}})(\psi_{K_{2}}^{\dagger}\sigma_{x}\psi_{K_{2}})$
where $\sigma_{x}$ is the pseudospin Pauli matrix along the $x$-axis.
Similarly, the out-of-plane arrows ($\odot$, $\otimes$) denote the
out-of-plane pseudospin moment represented by $\sigma_{z}$. The black
dots imply that the interaction can be regarded as a density-density
interaction. Note that the density operator is represented by $\sigma_{0}$. 

\begin{figure}
\includegraphics[width=0.9\textwidth]{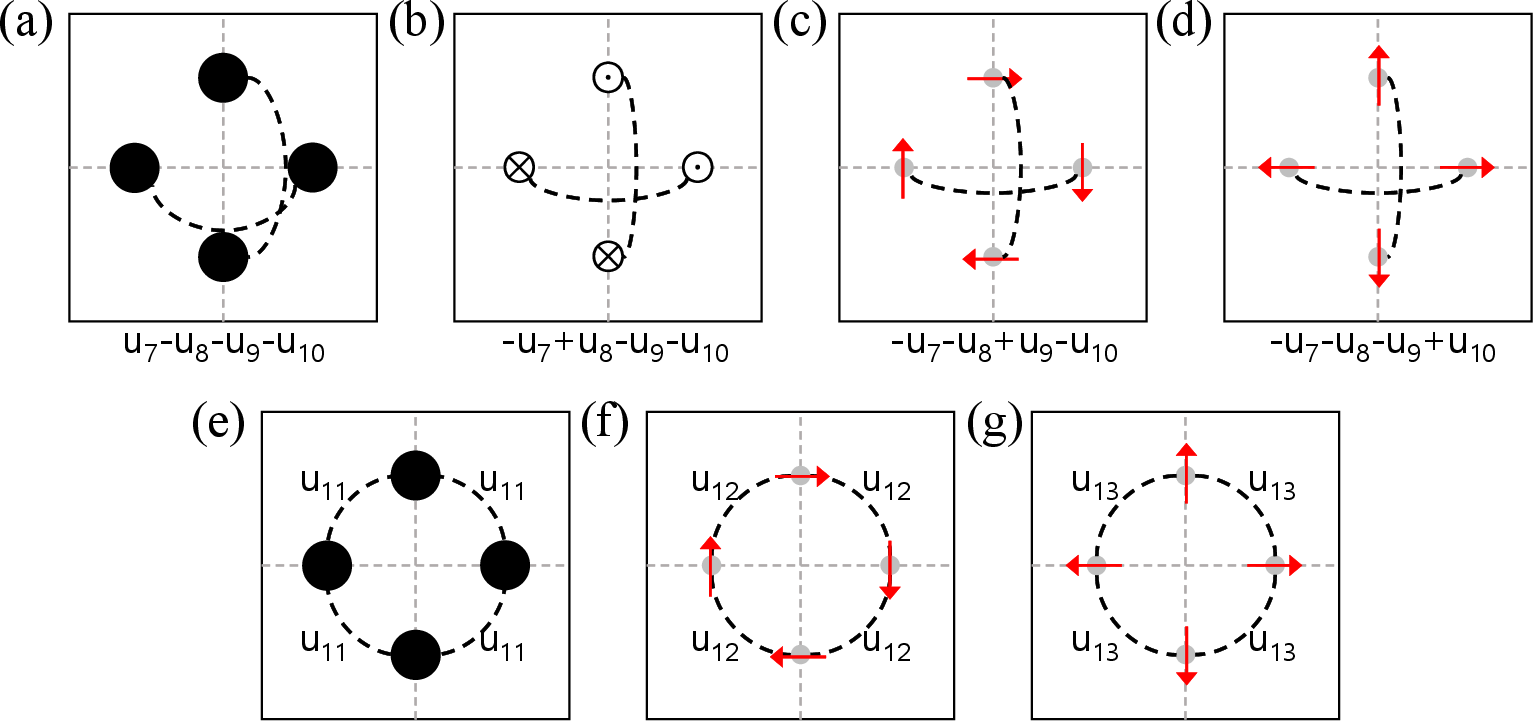}\caption{\label{fig:S2}Description of the interactions $u_{7}$-$u_{10}$
in the square lattice in the particle-particle channel by means of
pseudospin moment. Patches are marked by small gray disks. (a) schematic
pictures for linear combinations of $u_{7}$-$u_{10}$ for spin-singlet
and spin-triplet pairings. (b) Inter-patch pair hopping interactions
$u_{11}$-$u_{13}$ deciding the difference between the signs of gap
functions at the patches. Black circles denote the spin-singlet Cooper
pairs. $\odot$ and arrows denote the alignment of the total pseudospin
of spin-triplet Cooper pairs. Dashed lines links the patches involved
in the scattering due to the associated interaction.}
\end{figure}

When it comes to the interactions $u_{11}$-$u_{13}$, these interactions
describe inter-patch pair-hopping interactions of Cooper pairs, and
thus they have direct relevance to Cooper pairing. Note that $u_{7}$-$u_{10}$
are intra-patch interactions supporting or suppressing Cooper pairing
in this point of view. Subsequently, the description of these interactions
in the particle-particle channel provides insightful picture, whose
expressions are given by 

\begin{align}
u_{7}: & \frac{1}{2}\sum_{i=1,2}\bigg[\psi_{K_{i}}^{\dagger}(\check{\sigma}_{0}\cdot i\sigma_{y})\psi_{\bar{K}_{i}}^{\dagger T}\cdot\psi_{\bar{K}_{i}}^{T}(\check{\sigma}_{0}\cdot i\sigma_{y})^{\dagger}\psi_{K_{i}}+\psi_{K_{i}}^{\dagger}(\check{\sigma}_{z}\cdot i\sigma_{y})\psi_{\bar{K}_{i}}^{\dagger T}\cdot\psi_{\bar{K}_{i}}^{T}(\check{\sigma}_{z}\cdot i\sigma_{y})^{\dagger}\psi_{K_{i}}\\
 & +\psi_{K_{i}}^{\dagger}(\check{\sigma}_{x}\cdot i\sigma_{y})\psi_{\bar{K}_{i}}^{\dagger T}\cdot\psi_{\bar{K}_{i}}^{T}(\check{\sigma}_{x}\cdot i\sigma_{y})^{\dagger}\psi_{K_{i}}+\psi_{K_{i}}^{\dagger}(\check{\sigma}_{y}\cdot i\sigma_{y})\psi_{\bar{K}_{i}}^{\dagger T}\cdot\psi_{\bar{K}_{i}}^{T}(\check{\sigma}_{y}\cdot i\sigma_{y})^{\dagger}\psi_{K_{i}}\bigg],\nonumber \\
u_{8}: & \frac{1}{2}\sum_{i=1,2}\bigg[-\psi_{K_{i}}^{\dagger}(\check{\sigma}_{0}\cdot i\sigma_{y})\psi_{\bar{K}_{i}}^{\dagger T}\cdot\psi_{\bar{K}_{i}}^{T}(\check{\sigma}_{0}\cdot i\sigma_{y})^{\dagger}\psi_{K_{i}}-\psi_{K_{i}}^{\dagger}(\check{\sigma}_{z}\cdot i\sigma_{y})\psi_{\bar{K}_{i}}^{\dagger T}\cdot\psi_{\bar{K}_{i}}^{T}(\check{\sigma}_{z}\cdot i\sigma_{y})^{\dagger}\psi_{K_{i}}\\
 & +\psi_{K_{i}}^{\dagger}(\check{\sigma}_{x}\cdot i\sigma_{y})\psi_{\bar{K}_{i}}^{\dagger T}\cdot\psi_{\bar{K}_{i}}^{T}(\check{\sigma}_{x}\cdot i\sigma_{y})^{\dagger}\psi_{K_{i}}+\psi_{K_{i}}^{\dagger}(\check{\sigma}_{y}\cdot i\sigma_{y})\psi_{\bar{K}_{i}}^{\dagger T}\cdot\psi_{\bar{K}_{i}}^{T}(\check{\sigma}_{y}\cdot i\sigma_{y})^{\dagger}\psi_{K_{i}}\bigg],\nonumber \\
u_{9}: & \frac{1}{2}\sum_{i=1,2}\bigg[-\psi_{K_{i}}^{\dagger}(\check{\sigma}_{0}\cdot i\sigma_{y})\psi_{\bar{K}_{i}}^{\dagger T}\cdot\psi_{\bar{K}_{i}}^{T}(\check{\sigma}_{0}\cdot i\sigma_{y})^{\dagger}\psi_{K_{i}}+\psi_{K_{i}}^{\dagger}(\check{\sigma}_{z}\cdot i\sigma_{y})\psi_{\bar{K}_{i}}^{\dagger T}\cdot\psi_{\bar{K}_{i}}^{T}(\check{\sigma}_{z}\cdot i\sigma_{y})^{\dagger}\psi_{K_{i}}\\
 & -\psi_{K_{i}}^{\dagger}(\check{\sigma}_{x}\cdot i\sigma_{y})\psi_{\bar{K}_{i}}^{\dagger T}\cdot\psi_{\bar{K}_{i}}^{T}(\check{\sigma}_{x}\cdot i\sigma_{y})^{\dagger}\psi_{K_{i}}+\psi_{K_{i}}^{\dagger}(\check{\sigma}_{y}\cdot i\sigma_{y})\psi_{\bar{K}_{i}}^{\dagger T}\cdot\psi_{\bar{K}_{i}}^{T}(\check{\sigma}_{y}\cdot i\sigma_{y})^{\dagger}\psi_{K_{i}}\bigg],\nonumber \\
u_{10}: & \frac{1}{2}\sum_{i=1,2}\bigg[-\psi_{K_{i}}^{\dagger}(\check{\sigma}_{0}\cdot i\sigma_{y})\psi_{\bar{K}_{i}}^{\dagger T}\cdot\psi_{\bar{K}_{i}}^{T}(\check{\sigma}_{0}\cdot i\sigma_{y})^{\dagger}\psi_{K_{i}}+\psi_{K_{i}}^{\dagger}(\check{\sigma}_{z}\cdot i\sigma_{y})\psi_{\bar{K}_{i}}^{\dagger T}\cdot\psi_{\bar{K}_{i}}^{T}(\check{\sigma}_{z}\cdot i\sigma_{y})^{\dagger}\psi_{K_{i}},\nonumber \\
 & +\psi_{K_{i}}^{\dagger}(\check{\sigma}_{x}\cdot i\sigma_{y})\psi_{\bar{K}_{i}}^{\dagger T}\cdot\psi_{\bar{K}_{i}}^{T}(\check{\sigma}_{x}\cdot i\sigma_{y})^{\dagger}\psi_{K_{i}}-\psi_{K_{i}}^{\dagger}(\check{\sigma}_{y}\cdot i\sigma_{y})\psi_{\bar{K}_{i}}^{\dagger T}\cdot\psi_{\bar{K}_{i}}^{T}(\check{\sigma}_{y}\cdot i\sigma_{y})^{\dagger}\psi_{K_{i}}\bigg],\\
u_{11}: & \psi_{K_{1}}^{\dagger}(\check{\sigma}_{0}\cdot i\sigma_{y})\psi_{\bar{K}_{1}}^{\dagger T}\cdot\psi_{\bar{K}_{2}}^{T}(\check{\sigma}_{0}\cdot i\sigma_{y})^{\dagger}\psi_{K_{2}}+{\rm h.c},\\
u_{12}: & \psi_{K_{1}}^{\dagger}(\check{\sigma}_{x}\cdot i\sigma_{y})\psi_{\bar{K}_{1}}^{\dagger T}\cdot\psi_{\bar{K}_{2}}^{T}(\check{\sigma}_{x}\cdot i\sigma_{y})^{\dagger}\psi_{K_{2}}+{\rm h.c},\\
u_{13}: & \psi_{K_{1}}^{\dagger}(\check{\sigma}_{y}\cdot i\sigma_{y})\psi_{\bar{K}_{1}}^{\dagger T}\cdot\psi_{\bar{K}_{2}}^{T}(\check{\sigma}_{y}\cdot i\sigma_{y})^{\dagger}\psi_{K_{2}}+{\rm h.c},\label{eq:u13}
\end{align}
with $\bar{K}_{i}=-K_{i}=K_{i+2}$. Here, $\check{\sigma}_{\mu}$
is defined by $\psi_{K_{i}}^{\dagger}\check{\sigma}_{\mu}(i\sigma_{y})\psi_{\bar{K}_{i}}^{\dagger T}=\psi_{K_{i}}^{\dagger}e^{-i\frac{\phi_{i}}{2}\sigma_{z}}\sigma_{\mu}e^{+i\frac{\phi_{i}}{2}\sigma_{z}}(i\sigma_{y})\psi_{\bar{K}_{i}}^{\dagger T}$
for each $i=1,2,3,4$, respectively. Note that terms of pair-hopping
interactions involving $\sigma_{z}$ such as $\psi_{K_{1}}^{\dagger}(\check{\sigma}_{z}\cdot i\sigma_{y})\psi_{\bar{K}_{1}}^{\dagger T}\cdot\psi_{\bar{K}_{2}}^{T}(\check{\sigma}_{z}\cdot i\sigma_{y})^{\dagger}\psi_{K_{2}}$
are forbidden by mirror symmetries. Note that linear combinations
$u_{7}$-$u_{10}$ correspond directly to the pairing interactions
for superconductivity
\begin{align}
u_{7}-u_{8}-u_{9}-u_{10}: & \frac{1}{2}\sum_{i=1,2}\psi_{K_{i}}^{\dagger}(\check{\sigma}_{0}\cdot i\sigma_{y})\psi_{\bar{K}_{i}}^{\dagger T}\cdot\psi_{\bar{K}_{i}}^{T}(\check{\sigma}_{0}\cdot i\sigma_{y})^{\dagger}\psi_{K_{i}},\\
-u_{7}+u_{8}-u_{9}-u_{10}: & \frac{1}{2}\sum_{i=1,2}\psi_{K_{i}}^{\dagger}(\check{\sigma}_{z}\cdot i\sigma_{y})\psi_{\bar{K}_{i}}^{\dagger T}\cdot\psi_{\bar{K}_{i}}^{T}(-\check{\sigma}_{z}\cdot i\sigma_{y})^{\dagger}\psi_{K_{i}},\\
-u_{7}-u_{8}+u_{9}-u_{10}: & \frac{1}{2}\sum_{i=1,2}\psi_{K_{i}}^{\dagger}(\check{\sigma}_{x}\cdot i\sigma_{y})\psi_{\bar{K}_{i}}^{\dagger T}\cdot\psi_{\bar{K}_{i}}^{T}(-\check{\sigma}_{x}\cdot i\sigma_{y})^{\dagger}\psi_{K_{i}},\\
-u_{7}-u_{8}-u_{9}+u_{10}: & \frac{1}{2}\sum_{i=1,2}\psi_{K_{i}}^{\dagger}(\check{\sigma}_{y}\cdot i\sigma_{y})\psi_{\bar{K}_{i}}^{\dagger T}\cdot\psi_{\bar{K}_{i}}^{T}(-\check{\sigma}_{y}\cdot i\sigma_{y})^{\dagger}\psi_{K_{i}}.
\end{align}

Figure \ref{fig:S2} displays these interaction in terms of the bilinears
in the particle-particle channel: $\psi_{K_{i}}^{\dagger}(\sigma_{\mu}i\sigma_{y})\psi_{\bar{K}_{i}}^{\dagger T}$
and $\psi_{\bar{K}_{i}}^{T}(\sigma_{\mu}i\sigma_{y})^{\dagger}\psi_{K_{i}}$
($\mu=0,x,y,z$). The black dot and red arrows represent the pseudospin-singlet
and the spin alignment of pseudospin-triplet pairings, respectively.
Note the pair-hopping interaction terms like $\psi_{K_{1}}^{\dagger}(\sigma_{x}i\sigma_{y})\psi_{\bar{K}_{1}}^{\dagger T}\cdot\psi_{\bar{K}_{2}}^{T}(\sigma_{x}i\sigma_{y})^{\dagger}\psi_{K_{2}}$
are forbidden by the point group symmetries. For example, under the
mirror reflection $M_{010}$ 
\begin{align}
M_{010}^{\dagger}\psi_{K_{1}}^{\dagger T}(\sigma_{x}i\sigma_{y})\psi_{\bar{K}_{1}}^{\dagger}M_{010}= & +\psi_{K_{1}}^{\dagger T}(\sigma_{x}i\sigma_{y})\psi_{\bar{K}_{1}}^{\dagger},\\
M_{010}^{\dagger}\psi_{\bar{K}_{2}}^{T}(\sigma_{x}i\sigma_{y})^{\dagger}\psi_{K_{2}}M_{010}= & -\psi_{\bar{K}_{2}}^{T}(\sigma_{x}i\sigma_{y})^{\dagger}\psi_{K_{2}},
\end{align}
and thus such an interaction is not invariant, however becomes available
in systems with lower symmetry group.

All interactions $u_{1}$-$u_{10}$ shown in Fig. \ref{fig:S1} are
able to trigger Pomeranchuk instabilities preserving the translational
symmetries. Density-wave type instabilities along with the broken
translational symmetries with nesting momentum $\boldsymbol{q}=\boldsymbol{Q}_{1}$
($\boldsymbol{q}=\boldsymbol{Q}_{2}$) are also caused by $u_{2}$-$u_{6}$
and $u_{11}$-$u_{13}$ ($u_{7}$-$u_{10}$).

We would like to note that the expression of interactions in Eqs.
\eqref{eq:u1}-\eqref{eq:u13} are derived solely by the symmetries
of the normal phase. Though we use $D_{{\rm 4h}}$-symmetric lattice,
the four-fold rotation symmetry $C_{{\rm 4}z}$ and two mirror symmetries
against the planes in which the VHSs are found are important in deriving
the interactions (alternatively, the two-fold rotation symmetries
along the axis normal to the planes can be used). Subsequently, it
can also be used in the square lattices subject to the point groups
such as $C_{{\rm 4h}}$ and $D_{{\rm 2h}}$ as long as (i) the pseudospin
basis is available, and (ii) there is just one star of type-II VHSs
(or four VHSs). Concerning the out-of-plane mirror symmetry, its major
role in the current model is disallowing the Ising type spin-orbit
coupling proportional to $\tau_{z}s_{z}$ in $H_{{\rm TB}}^{({\rm LNCS})}$.
However, even though this SOC is allowed, the eigenstates still form
the pseudospin basis, and thus the form of the interactions given
in Eqs. \eqref{eq:u1}-\eqref{eq:u13} hold true.

\subsection{RG equations for the coupling constants $u_{i}$}

In the derivation of the RG equations for the coupling constant, we
need to evaluate the logarithmically divergent one-loop diagrams.
Such divergences are characterized by the particle-particle bare polarization
function $\Pi_{\boldsymbol{q}}^{({\rm pp})}(\Omega)=\sum_{\omega,\boldsymbol{k}}\text{Tr}[G(\boldsymbol{k},\omega)G(\boldsymbol{q}-\boldsymbol{k},\Omega-\omega)]$
and particle-hole bare polarization function $\Pi_{\boldsymbol{q}}^{({\rm ph})}(\Omega)=\sum_{\omega,\boldsymbol{k}}\text{Tr}[G(\boldsymbol{k},\omega)G(\boldsymbol{q}+\boldsymbol{k},\Omega+\omega)]$.
Here, $\boldsymbol{q}$ and $\Omega$ denote the total momentum and
the frequency of the pair of particles in $\Pi_{\boldsymbol{q}}^{({\rm pp})}(\Omega)$.
When it comes to $\Pi_{\boldsymbol{q}}^{({\rm ph})}(\Omega)$, they
mean the momentum and the frequency transferred between the particle-hole
pairs scattered through an interaction. In the patch approximation
to the whole electronic structure, the nesting between VHSs are important
and the susceptibilities evaluated at $\boldsymbol{q}=\boldsymbol{Q}_{i}$
are of the importance, where $\boldsymbol{Q}_{i}$ is a nesting momentum
relating two patches. When $\Omega$ is low, the density of states
begins to exhibit its characteristic logarithmic divergence around
the VHSs and the six relevant particle-particle and particle-hole
bare polarization functions are given by 
\begin{align}
\Pi_{\boldsymbol{0}}^{({\rm pp})}(\Omega)\approx\nu_{0}\ln^{2}\frac{\Lambda}{\Omega},\quad\:\; & \Pi_{\boldsymbol{0}}^{({\rm ph})}(\Omega)\approx2\nu_{0}\ln\frac{\Lambda}{\Omega},\\
\Pi_{\boldsymbol{Q}_{1}}^{({\rm pp})}(\Omega)\approx2a_{1}\nu_{0}\ln\frac{\Lambda}{\Omega},\,\,\, & \Pi_{\boldsymbol{Q}_{1}}^{({\rm ph})}(\Omega)\approx2\beta_{1}\nu_{0}\ln\frac{\Lambda}{\Omega},\\
\Pi_{\boldsymbol{Q}_{2}}^{({\rm pp})}(\Omega)\approx a_{2}\nu_{0}\ln^{2}\frac{\Lambda}{\Omega},\,\,\, & \Pi_{\boldsymbol{Q}_{2}}^{({\rm ph})}(\Omega)\approx2a_{2}\nu_{0}\ln\frac{\Lambda}{\Omega},
\end{align}
with $\nu_{0}=\frac{\sqrt{m_{x}m_{y}}}{4\pi^{2}}$ a measure of the
density of states around VHSs. Here, $\Lambda$ denotes the ultraviolet
cutoff up to which the current patch model remains valid. $\Omega$
plays the role of the infrared cutoff which can be replaced with temperature
$T\neq0$. An analytical estimation of the parameters $a_{1}$ and
$\beta_{1}$ yields $a_{1}=\frac{1+\kappa}{2\sqrt{\kappa}}$ and $\beta_{1}=\frac{2\sqrt{\kappa}}{1+\kappa}\ln|\frac{\kappa+1}{\kappa-1}|$
where $\kappa=m_{y}/m_{x}$ by assuming the quadratic energy dispersion
$\varepsilon(\boldsymbol{k})=\frac{1}{2m_{x}}k_{x}^{2}-\frac{1}{2m_{y}}k_{y}^{2}$
around VHS at $K_{1}$. Though $a_{2}=1$ in the same quadratic approximation
of the energy dispersion, terms higher order in $\boldsymbol{k}$
in the energy dispersion spoil the nesting, and thus we let it variable
between 0 and 1.

Given $\Pi_{\boldsymbol{q}}^{({\rm pp})}$ and $\Pi_{\boldsymbol{q}}^{({\rm ph})}$,
the RG equations for the coupling constants are given by

\begin{align}
\dot{u}_{1}= & d_{0}^{{\rm (ph)}}\left(2u_{1}^{2}+u_{10}^{2}-2u_{2}^{2}+2u_{3}^{2}+2u_{4}^{2}+4u_{5}^{2}+2u_{6}^{2}-u_{7}^{2}+u_{8}^{2}+u_{9}^{2}\right)-2d_{1}^{{\rm (pp)}}u_{1}^{2},\label{eq:RGeq_u1}\\
\dot{u}_{2}= & -4d_{0}^{{\rm (ph)}}u_{2}(u_{1}+u_{7})+d_{1}^{{\rm (ph)}}\left(u_{2}^{2}+u_{3}^{2}+u_{4}^{2}+2u_{5}^{2}+u_{6}^{2}+u_{11}^{2}+u_{12}^{2}+u_{13}^{2}\right)\\
 & -d_{2}^{{\rm {\rm (pp)}}}\left(u_{2}^{2}+u_{3}^{2}+u_{4}^{2}+2u_{5}^{2}+u_{6}^{2}\right),\nonumber \\
\dot{u}_{3}= & 4d_{0}^{{\rm (ph)}}u_{3}(u_{1}-u_{8})+2d_{1}^{{\rm (ph)}}\left(u_{12}u_{13}+u_{2}u_{3}+u_{4}u_{6}+u_{5}^{2}\right)+2d_{2}^{{\rm {\rm (pp)}}}\left(-u_{2}u_{3}+u_{4}u_{6}+u_{5}^{2}\right),\\
\dot{u}_{4}= & 4d_{0}^{{\rm (ph)}}u_{4}(u_{1}+u_{9})+2d_{1}^{{\rm (ph)}}(u_{11}u_{12}+u_{2}u_{4}+u_{3}u_{6})+2d_{2}^{{\rm {\rm (pp)}}}(u_{3}u_{6}-u_{2}u_{4}),\\
\dot{u}_{5}= & 2d_{0}^{{\rm (ph)}}u_{5}(2u_{1}-u_{10}-u_{9})+2d_{1}^{{\rm (ph)}}u_{5}(u_{2}+u_{3})+2d_{2}^{{\rm {\rm (pp)}}}u_{5}(u_{3}-u_{2}),\\
\dot{u}_{6}= & 4d_{0}^{{\rm (ph)}}u_{6}(u_{1}+u_{10})+2d_{1}^{{\rm (ph)}}(u_{11}u_{13}+u_{2}u_{6}+u_{3}u_{4})+2d_{2}^{{\rm {\rm (pp)}}}(u_{3}u_{4}-u_{2}u_{6}),\\
\dot{u}_{7}= & -4d_{0}^{{\rm (ph)}}\left(u_{1}u_{7}+u_{2}^{2}\right)+d_{2}^{{\rm (ph)}}\left(u_{10}^{2}+u_{7}^{2}+u_{8}^{2}+u_{9}^{2}\right)-\left(u_{7}^{2}+u_{8}^{2}+u_{9}^{2}+u_{10}^{2}+u_{11}^{2}+u_{12}^{2}+u_{13}^{2}\right),\\
\dot{u}_{8}= & -4d_{0}^{{\rm (ph)}}\left(u_{3}^{2}-u_{1}u_{8}\right)+2d_{2}^{{\rm (ph)}}(u_{10}u_{9}+u_{7}u_{8})+2u_{10}u_{9}+u_{11}^{2}-u_{12}^{2}-u_{13}^{2}-2u_{7}u_{8},\\
\dot{u}_{9}= & 4d_{0}^{{\rm (ph)}}\left(u_{1}u_{9}+u_{4}^{2}-u_{5}^{2}\right)+d_{2}^{{\rm (ph)}}(2u_{10}u_{8}+2u_{7}u_{9})+2u_{10}u_{8}+u_{11}^{2}+u_{12}^{2}-u_{13}^{2}-2u_{7}u_{9},\\
\dot{u}_{10}= & 4d_{0}^{{\rm (ph)}}\left(u_{1}u_{10}-u_{5}^{2}+u_{6}^{2}\right)+2d_{2}^{{\rm (ph)}}(u_{10}u_{7}+u_{8}u_{9})-2u_{10}u_{7}+u_{11}^{2}-u_{12}^{2}+u_{13}^{2}+2u_{8}u_{9},\\
\dot{u}_{11}= & 4d_{1}^{{\rm (ph)}}(u_{11}u_{2}+u_{12}u_{4}+u_{13}u_{6})+2u_{11}(-u_{7}+u_{8}+u_{9}+u_{10}),\\
\dot{u}_{12}= & 4d_{1}^{{\rm (ph)}}(u_{11}u_{4}+u_{12}u_{2}+u_{13}u_{3})+2u_{12}(-u_{7}-u_{8}+u_{9}-u_{10}),\\
\dot{u}_{13}= & 4d_{1}^{{\rm (ph)}}(u_{11}u_{6}+u_{12}u_{3}+u_{13}u_{2})+2u_{13}(-u_{7}-u_{8}-u_{9}+u_{10}),\label{eq:RGeq_u13}
\end{align}
where $\dot{u}_{i}=\partial_{y}u_{i}(y)$ with $g_{i}=\nu_{0}u_{i}$
and the running scale $y=\Pi_{0}^{{\rm pp}}(\Omega)/\nu_{0}$. $d_{i}^{({\rm pp})}(y)=\partial_{y}\Pi_{\boldsymbol{Q}_{i}}^{({\rm pp})}/\nu_{0}$
and $d_{i}^{({\rm ph})}(y)=-\partial_{y}\Pi_{\boldsymbol{Q}_{i}}^{({\rm ph})}/\nu_{0}$.
$d_{i}^{({\rm pp}/{\rm ph})}(y)$ are generally decreasing function
of $y$ since $\Pi_{Q_{i}}^{({\rm pp}/{\rm ph})}$ usually logarithmically
divergent at low $\Omega$. Hence, we employ decreasing functions
$d_{2}^{({\rm pp})}(y)=\frac{1+a_{2}y}{1+y}$ and $d_{i}^{({\rm pp}/{\rm ph})}(y)=\frac{a_{i}}{\sqrt{a_{i}^{2}+y}}$
for the remaining to interpolate both limits of $d_{i}^{(c)}(y)$:
$\lim_{y\rightarrow0}d_{i}^{(c)}(y)=1$ and $\lim_{y\rightarrow\infty}d_{i}^{(c)}(0)=a_{i}/\sqrt{y}$
\citep{Furukawa1998,Hur2009}.

The microscopic details of the normal phase enters into this RG analysis
through the bare values $u_{i}(0)$ of the coupling constants $u_{i}$.
$u_{i}(0)$ are obtained by projecting, for example, the bare Hubbard
onto the pseudospin basis relevant to the VHSs around the Fermi level.
The resultant $u_{i}(0)$ are listed in Table \ref{tab:1} where $u_{i}(0)$
in the limit of no spin-orbit coupling or no interlayer hopping are
also shown together. 

\begin{table}
\begin{tabular*}{0.95\textwidth}{@{\extracolsep{\fill}}|c|>{\centering}m{0.045\textwidth}|>{\centering}m{0.045\textwidth}|>{\centering}m{0.05\textwidth}|>{\centering}m{0.045\textwidth}|>{\centering}m{0.045\textwidth}|>{\centering}m{0.045\textwidth}|>{\centering}m{0.07\textwidth}|>{\centering}m{0.05\textwidth}|>{\centering}m{0.05\textwidth}|>{\centering}m{0.09\textwidth}|>{\centering}m{0.07\textwidth}|>{\centering}m{0.05\textwidth}|>{\centering}m{0.05\textwidth}|}
\hline 
\noalign{\vskip\doublerulesep}
 & $u_{1}$ & $u_{2}$ & $u_{3}$ & $u_{4}$ & $u_{5}$ & $u_{6}$ & $u_{7}$ & $u_{8}$ & $u_{9}$ & $u_{10}$ & $u_{11}$ & $u_{12}$ & $u_{13}$\tabularnewline[\doublerulesep]
\hline 
\hline 
\noalign{\vskip\doublerulesep}
General case & \noindent \centering{}$\frac{U}{4}c^{2}$ & \noindent \centering{}$\frac{U}{4}$ & \noindent \centering{}$-\frac{U}{4}c^{2}$ & \noindent \centering{}$\frac{U}{4}s^{2}$ & \noindent \centering{}$-\frac{U}{4}c$ & \noindent \centering{}0 & \noindent \centering{}$\frac{U}{4}(1+s^{2})$ & \noindent \centering{}$-\frac{U}{4}c^{2}$ & \noindent \centering{}$-\frac{U}{4}s^{2}$ & \noindent \centering{}$-\frac{U}{4}c^{2}$ & \noindent \centering{}$\frac{U}{2}$ & \noindent \centering{}$\frac{U}{2}s^{2}$ & \noindent \centering{}0\tabularnewline[\doublerulesep]
\hline 
\noalign{\vskip\doublerulesep}
$\boldsymbol{\alpha}_{\boldsymbol{k}}|_{{\rm VHS}}=0$ & \noindent \centering{}$\frac{U}{4}$ & \noindent \centering{}$\frac{U}{4}$ & \noindent \centering{}$-\frac{U}{4}$ & \noindent \centering{}0 & \noindent \centering{}$-\frac{U}{4}$ & \noindent \centering{}0 & \noindent \centering{}$\frac{U}{4}$ & \noindent \centering{}$-\frac{U}{4}$ & \noindent \centering{}0 & \noindent \centering{}$-\frac{U}{4}$ & \noindent \centering{}$\frac{U}{2}$ & \noindent \centering{}0 & \noindent \centering{}0\tabularnewline[\doublerulesep]
\hline 
\noalign{\vskip\doublerulesep}
$\varepsilon_{x,\boldsymbol{k}}|_{{\rm VHS}}=0$ & \noindent \centering{}0 & \noindent \centering{}$\frac{U}{4}$ & \noindent \centering{}0 & \noindent \centering{}$\frac{U}{4}$ & \noindent \centering{}0 & \noindent \centering{}0 & \noindent \centering{}$\frac{U}{2}$ & \noindent \centering{}0 & \noindent \centering{}$-\frac{U}{4}$ & \noindent \centering{}0 & \noindent \centering{}$\frac{U}{2}$ & \noindent \centering{}$\frac{U}{2}$ & \noindent \centering{}0\tabularnewline[\doublerulesep]
\hline 
\end{tabular*}\caption{\label{tab:1}The bare values of $u_{i}(0)$. $c=\cos\zeta|_{\text{VHS}}$
and $s=\sin\zeta|_{\text{VHS}}$. The first row show $u_{i}(0)$ for
general situation. The second (third) row shows the case where the
effective spin-orbit coupling (interlayer hopping) at the VHSs is
negligible.}
\end{table}

\section{Cooper pairing and Density wave Instabilities of Normal Phase}

As explained in Sec. \ref{sec:Interactions}, there are three kinds
of nesting momentum: $\boldsymbol{0},\boldsymbol{Q}_{1},\boldsymbol{Q}_{2}$.
When it comes to the particle-particle (particle-hole) channels, the
nesting momentum are related with the total momentum (transferred
momentum) of the particle-particle pair (particle-hole pair). The
Cooper pairing (Pomeranchuk) channels involves a zero total (transferred)
momentum, and thus they can be regarded as particle-particle (particle-hole)
channels with $\boldsymbol{Q}=\boldsymbol{0}$. The pair-density wave
type particle-particle channels, which can trigger the FFLO-type superconductivity
or pair-density wave superconductivity, involves finite $\boldsymbol{Q}$.
The density-wave channels are particle-hole channels with finite transferred
momentum. Thus, the whole instability channels can be classified into
three particle-particle channels and three particle-hole channels.
However, the Cooper pairing channels and the density wave channels
with $\boldsymbol{Q}=\boldsymbol{Q}_{1}$ are most relevant to our
result, and thus this section will be devoted to these two channels.

\subsection{Cooper pairing instabilities}

\begin{figure}
\includegraphics[width=0.8\textwidth]{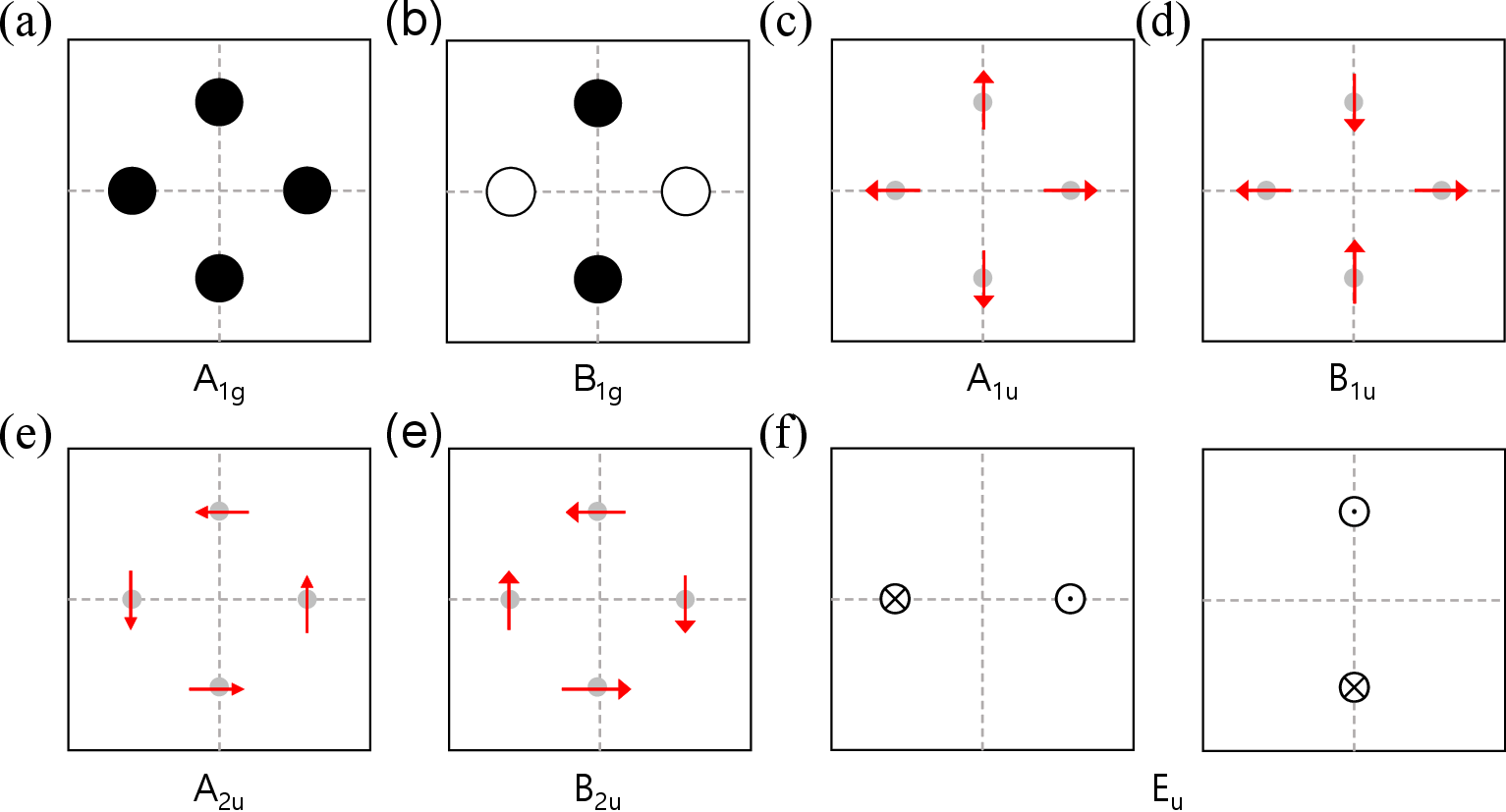}\caption{\label{fig:S3}
Pseudospin representation of superconducting gap functions
allowed in the current patch approximation. Patches are marked by
small gray disks. In (b), the empty disks represent the spin-singlet
gap function with the sign opposite to that of the filled black disks.
In (f), the two degenerate gap functions of irrep $E_{{\rm u}}$ are
depicted separately.}
\end{figure}

The instability of the normal phase can be examined by inspecting
the susceptibility $\chi$ of the system against auxiliary probing
vertices $\Delta$. To do this, we first need to introduce the auxiliary
probing vertices $\Delta$ to our Hamiltonian and then obtain the
RG equations for them. As an example, let us consider the Cooper pairing
instabilities. For this, we add the following bilnears to our Hamiltonian
\begin{equation}
V_{0}^{({\rm pp})}=\frac{1}{2}\sum_{m}\bigg[\Delta_{\boldsymbol{0},m}^{({\rm pp)}}\sum_{\boldsymbol{k}}\psi^{\dagger}(\boldsymbol{k})M_{\boldsymbol{0},m}^{({\rm pp})}\psi^{\dagger T}(-\boldsymbol{k})+\Delta_{\boldsymbol{0},m}^{({\rm pp)}*}\sum_{\boldsymbol{k}}\psi^{T}(-\boldsymbol{k})[M_{\boldsymbol{0},m}^{({\rm pp})}]^{\dagger}\psi(\boldsymbol{k})\bigg],
\end{equation}
with $\psi(\boldsymbol{k})=(\psi_{\boldsymbol{K}_{1}+\boldsymbol{k}},\psi_{\boldsymbol{K}_{2}+\boldsymbol{k}},\psi_{\boldsymbol{K}_{3}+\boldsymbol{k}},\psi_{\boldsymbol{K}_{4}+\boldsymbol{k}})^{T}$.
Here, $m$ can run over the irreducible representations of the point
group of the normal phase. The complex numbers $\Delta_{\boldsymbol{0},m}^{({\rm pp)}}$
and $\Delta_{\boldsymbol{0},m}^{({\rm pp)}*}$ measure the amplitude
of the perturbation triggering Cooper pairing while matrices $M_{\boldsymbol{0},m}^{({\rm pp})}$
describe the form factor of the Cooper pairs. Expressed in the matrix
form, $M_{\boldsymbol{0},m}^{({\rm pp})}$ can be written as
\begin{equation}
M_{\boldsymbol{0},m}^{({\rm pp})}=\left(\begin{matrix} &  & m_{1\bar{1}}\\
 &  &  & m_{2\bar{2}}\\
-m_{1\bar{1}}^{T}\\
 & -m_{2\bar{2}}^{T}
\end{matrix}\right),
\end{equation}
where $m_{i\bar{i}}$ are $2\times2$ matrices. Other terms not appearing
in $M_{\boldsymbol{0},m}^{({\rm pp})}$ are associated to the pairings
of electrons with finite total momentum.

Among the ten irreducible representations (irreps) of the point group
$D_{{\rm 4h}}$, the Cooper pairings belonging to two one-dimensional
(1d) irreps $A_{{\rm 2g}}$ and $B_{{\rm 2g}}$ can not be captured
in the current patch model due to the location of the patches. Due
to the constraints set by the symmetries, $m_{1\bar{1}}=m_{2\bar{2}}=0$
for $M_{0,A_{2g}}^{({\rm pp})}$ and $M_{0,B_{2g}}^{({\rm pp})}$.
Cooper pairing belong to the even-parity two-dimensional (2d) irrep
$E_{g}$ is also unavailable in two-dimensional lattices. The pseudospin
representation of the remaining six one-dimensional irreps ($A_{{\rm 1g}}$,
$B_{1g}$, $A_{{\rm 1u}}$, $B_{1u}$, $A_{{\rm 2u}}$, $B_{{\rm 2u}}$)
and the odd-parity two-dimensional irrep ($E_{u}$) are given in Figure
\ref{fig:S3}. The meaning of black dots, in-plane arrows, and the
out-of-plane arrows are the same as in Fig. \ref{fig:S2} and they
specify the right-upper block of $M_{0,m}^{({\rm pp})}$ as follows:
\begin{equation}
M_{\boldsymbol{0},A_{{\rm 1g}}}^{({\rm pp})}=\left(\begin{matrix}\check{\sigma}_{0}\\
 & \check{\sigma}_{0}
\end{matrix}\right)i\sigma_{y},\quad M_{\boldsymbol{0},B_{{\rm 1g}}}^{({\rm pp})}=\left(\begin{matrix}\check{\sigma}_{0}\\
 & -\check{\sigma}_{0}
\end{matrix}\right)i\sigma_{y},\quad M_{\boldsymbol{0},A_{{\rm 1u}}}^{({\rm pp})}=\left(\begin{matrix}\check{\sigma}_{y}\\
 & \check{\sigma}_{y}
\end{matrix}\right)i\sigma_{y},\quad M_{\boldsymbol{0},B_{{\rm 1u}}}^{({\rm pp})}=\left(\begin{matrix}-\check{\sigma}_{y}\\
 & \check{\sigma}_{y}
\end{matrix}\right)i\sigma_{y},
\end{equation}
\begin{equation}
M_{\boldsymbol{0},A_{{\rm 2u}}}^{({\rm pp})}=\left(\begin{matrix}\check{\sigma}_{x}\\
 & \check{\sigma}_{x}
\end{matrix}\right)i\sigma_{y},\quad M_{\boldsymbol{0},B_{{\rm 2u}}}^{({\rm pp})}=\left(\begin{matrix}\check{\sigma}_{x}\\
 & -\check{\sigma}_{x}
\end{matrix}\right)i\sigma_{y},\quad M_{\boldsymbol{0},E_{u}(a)}^{({\rm pp})}=\left(\begin{matrix}\check{\sigma}_{z}\\
 & 0
\end{matrix}\right)i\sigma_{y},\quad M_{\boldsymbol{0},E_{u}(b)}^{({\rm pp})}=\left(\begin{matrix}0\\
 & \check{\sigma}_{z}
\end{matrix}\right)i\sigma_{y},
\end{equation}
where the right hand side of each equality correspond to the right-upper
block of $M_{\boldsymbol{0},m}^{({\rm pp})}$.

The RG equations for $\Delta_{\boldsymbol{0},m}^{({\rm pp})}(y)$
can be obtained by calculating the one-loop corrections to $\Delta_{\boldsymbol{0},m}^{({\rm pp})}(y)$
from the interactions. Writing our interation as $\hat{V}=\sum_{i=1,\cdots,13}u_{i}\sum_{m}c_{i,a}(\psi^{\dagger}[M_{i,a}]\psi)(\psi^{\dagger}[N_{i,a}]^{\dagger}\psi)$,
the one-loop correction for $\Delta_{\boldsymbol{0},m}^{({\rm pp})}(y)$
is expressed as $\frac{\partial\Delta_{\boldsymbol{0},m}^{({\rm pp})}(y)}{\partial y}=[\Gamma_{\boldsymbol{0}}^{({\rm pp})}(\boldsymbol{u})]_{mm}\Delta_{\boldsymbol{0},m}^{({\rm pp})}(y)\frac{\partial}{\partial y}\bigg[\sum_{\omega,\boldsymbol{k}}G_{K_{i}}(k)G_{\bar{K}_{i}}(-k)\bigg]$
with $y=\Pi_{\boldsymbol{0}}^{({\rm pp})}(\Omega)/\nu_{0}$, $\boldsymbol{u}=(u_{1}(y),u_{2}(y),\cdots,u_{13}(y))$,
and 
\begin{align}
[\Gamma_{\boldsymbol{0}}^{({\rm pp})}(\boldsymbol{u})]_{mn}\equiv & \frac{2}{{\rm Tr}[M_{\boldsymbol{0},m}^{({\rm pp})\dagger}M_{\boldsymbol{0},m}^{({\rm pp})}]}\sum_{l=1,2}\sum_{i,a}u_{i}c_{i,a} \label{Gamma_pp} \\
 & \times\bigg[\frac{\partial^{2}\text{Tr}[M_{\boldsymbol{0},m}^{({\rm pp})\dagger}M_{i,a}G(k)M_{\boldsymbol{0},n}^{({\rm pp})}G(-k)N_{i,a}^{T}]}{\partial G_{K_{l}}(k)\partial G_{\bar{K}_{l}}(-k)}+\frac{\partial^{2}\text{Tr}[M_{\boldsymbol{0},m}^{({\rm pp})\dagger}N_{i,a}^{T}G(k)M_{\boldsymbol{0},n}^{({\rm pp})}G(-k)M_{i,a}]}{\partial G_{K_{l}}(k)\partial G_{\bar{K}_{l}}(-k)}\bigg].\nonumber 
\end{align}
Note that $\frac{\partial}{\partial y}\bigg[\sum_{\omega,\boldsymbol{k}}G_{K_{i}}(k)G_{\bar{K}_{i}}(-k)\bigg]=1$
for this Cooper pairing case. $\Gamma_{\boldsymbol{0},m}^{({\rm pp})}(\boldsymbol{u})$
for each Cooper pairing vertex are given by 
\begin{align}
\Gamma_{\boldsymbol{0},A_{{\rm 1g}}}^{({\rm pp})}(\boldsymbol{u})= & -u_{7}+u_{8}+u_{9}+u_{10}-2u_{11},\\
\Gamma_{\boldsymbol{0},B_{{\rm 1g}}}^{({\rm pp})}(\boldsymbol{u})= & -u_{7}+u_{8}+u_{9}+u_{10}+2u_{11},\\
\Gamma_{\boldsymbol{0},A_{{\rm 1u}}}^{({\rm pp})}(\boldsymbol{u})= & -u_{7}-u_{8}-u_{9}+u_{10}-2u_{13},\\
\Gamma_{\boldsymbol{0},B_{{\rm 1u}}}^{({\rm pp})}(\boldsymbol{u})= & -u_{7}-u_{8}-u_{9}+u_{10}+2u_{13},\\
\Gamma_{\boldsymbol{0},A_{{\rm 2u}}}^{({\rm pp})}(\boldsymbol{u})= & -u_{7}-u_{8}+u_{9}-u_{10}-2u_{13},\\
\Gamma_{\boldsymbol{0},B_{{\rm 2u}}}^{({\rm pp})}(\boldsymbol{u})= & -u_{7}-u_{8}+u_{9}-u_{10}+2u_{13},\\
\Gamma_{\boldsymbol{0},E_{{\rm u}}}^{({\rm pp})}(\boldsymbol{u})= & -u_{7}+u_{8}-u_{9}-u_{10}.
\end{align}

The susceptibility $\chi_{0,m}^{({\rm pp})}$ with respect to the
perturbing vertex $\Delta_{0,m}^{({\rm pp})}(y)$ is introduced to
the total Hamiltonian by adding $\frac{1}{2}|\Delta_{0,m}^{({\rm pp})}(0)|^{2}\chi_{0,m}^{({\rm pp})}(y)$.
Through this term, the contribution to the free energy from the electronic
states in the momentum shell which are integrated out during the RG
procedure is stored \citep{Nelson1975}. RG equation for $\chi_{0,m}^{({\rm pp})}$
can be obtained by contracting two diagrams representing $\Delta_{\boldsymbol{0},m}^{({\rm pp)}}\sum_{\boldsymbol{k}}\psi^{\dagger}(\boldsymbol{k})M_{\boldsymbol{0},m}^{({\rm pp})}\psi^{\dagger T}(-\boldsymbol{k})$
and $\Delta_{\boldsymbol{0},m}^{({\rm pp)}*}\sum_{\boldsymbol{k}}\psi^{T}(-\boldsymbol{k})[M_{\boldsymbol{0},m}^{({\rm pp})}]^{\dagger}\psi(\boldsymbol{k})$.
Through this procedure, we obtain
\begin{align}
\frac{\partial\chi_{0,m}^{({\rm pp})}(y)}{\partial y}= & \bigg|\frac{\Delta_{0,m}^{({\rm pp})}(y)}{\Delta_{0,m}^{({\rm pp})}(0)}\bigg|^{2}\sum_{l=1,2}\frac{\partial^{2}\text{Tr}[G(k)M_{\boldsymbol{0},m}^{({\rm pp})}[G(-k)]^{T}[M_{\boldsymbol{0},m}^{({\rm pp})}]^{\dagger}]}{\partial G_{K_{i}}(\boldsymbol{k})\partial G_{\bar{K}_{i}}(-\boldsymbol{k})}\times\frac{\partial}{\partial y}\bigg[\sum_{\omega,\boldsymbol{k}}G_{K_{i}}(k)G_{\bar{K}_{i}}(-k)\bigg]
\end{align}
which should be solved with $\chi_{0,m}^{({\rm pp})}(0)=0$. Here,
we use $M_{\boldsymbol{0},m}^{({\rm pp})}=-[M_{\boldsymbol{0},m}^{({\rm pp})}]^{T}$
derived from $M_{\boldsymbol{0},m}^{({\rm pp})}(-\boldsymbol{k})=-[M_{\boldsymbol{0},m}^{({\rm pp})}(\boldsymbol{k})]^{T}$.
Note that if $\Delta_{0,m}^{({\rm pp})}(y)$ diverges, $\chi_{0,m}^{({\rm pp})}(y)$
may diverge while the other factors in $\partial_{y}\chi_{0,m}^{({\rm pp})}(y)$
other than $|\Delta_{0,m}^{({\rm pp})}(y)|^{2}$ is not decisive when
it comes to the divergence of $\chi_{0,m}^{({\rm pp})}(y)$. The Cooper
pairing channel with the most largest $\chi_{0,m}^{({\rm pp})}(y)$
is the channel that the normal phase Fermi liquid is mostly prone
to transit into.

The asymptotic behavior of $\Delta_{\boldsymbol{0}}^{({\rm pp})}$
and $\chi_{0}^{({\rm pp})}$ can be guessed by noting that the asymptotic
behavior of the solution $u_{i}(y)$ of the RG equation in Eqs. \eqref{eq:RGeq_u1}-\eqref{eq:RGeq_u13}
is 
\begin{equation}
u_{i}(y)\sim u_{i}^{*}/(y_{c}-y)
\end{equation}
 for $y\approx y_{c}$. Due to this asymptotic behavior of the coupling
constants, the vertices $\Delta_{\boldsymbol{0},m}^{({\rm pp})}$
also diverge like $1/(y_{c}-y)^{\Gamma_{0,m}^{({\rm pp)}}(\boldsymbol{u}^{*})}$
which leads to $\chi_{0,m}^{({\rm pp})}\sim(y_{c}-y)^{-\alpha}$ with
the exponent 
\begin{equation}
\alpha=2\Gamma_{0,m}^{({\rm pp)}}(\boldsymbol{u}^{*})-1.\label{eq:exponent_SC}
\end{equation}
 Therefore, the channel with the most positive $\Gamma_{0,m}^{({\rm pp)}}(\boldsymbol{u}^{*})>\frac{1}{2}$
can be the leading instability of the normal phase in the weak-coupling
limit \citep{Lin2020,Maiti2010,Maiti2013}. 

\subsection{Density-wave instabilities}
\label{subsec:DW}
To examine the instabilities toward density wave phases, the auxiliary
probing vertices $\Delta$ for the denstiy-wave instabilities should
be added to our Hamiltonian. 

\begin{align}
V_{\boldsymbol{Q}_{1}}^{({\rm ph})}= & \frac{1}{2}\sum_{m}\bigg[\Delta_{\boldsymbol{Q}_{1},m}^{({\rm ph)}}\sum_{\boldsymbol{k}}\psi^{\dagger}(\boldsymbol{k})M_{\boldsymbol{Q}_{1},m}^{({\rm ph})}\psi(\boldsymbol{k})+\Delta_{\bar{\boldsymbol{Q}}_{1},m}^{({\rm ph)}}\sum_{\boldsymbol{k}}\psi^{\dagger}(\boldsymbol{k})M_{\bar{\boldsymbol{Q}}_{1},m}^{({\rm ph})}\psi(\boldsymbol{k})\bigg]\label{eq:V_DW}
\end{align}
where $\boldsymbol{Q}_{1}=(k,k)$ and $\bar{\boldsymbol{Q}}_{1}=(-k,k)$
with $0<k<\pi$. The derivation of this expression is presented in
Sec. \ref{subsec:derivation}.

Considering $\psi(\boldsymbol{k})=(\psi_{\boldsymbol{K}_{1}+\boldsymbol{k}},\psi_{\boldsymbol{K}_{2}+\boldsymbol{k}},\psi_{\boldsymbol{K}_{3}+\boldsymbol{k}},\psi_{\boldsymbol{K}_{4}+\boldsymbol{k}})^{T}$,
the matrix representations of $M_{\boldsymbol{Q},m}^{({\rm ph})}$
($\boldsymbol{Q}=\boldsymbol{Q}_{1},\bar{\boldsymbol{Q}}_{1}$) are
given by
\begin{equation}
M_{\boldsymbol{Q}_{1},m}^{({\rm ph})}=\left(\begin{matrix} &  &  & m_{14}\\
 &  & m_{23}\\
 & m_{32}\\
m_{41}
\end{matrix}\right),\quad M_{\bar{\boldsymbol{Q}}_{1},m}^{({\rm ph})}=\left(\begin{matrix} & m_{12}\\
m_{21}\\
 &  &  & m_{34}\\
 &  & m_{43}
\end{matrix}\right),
\end{equation}
respectively. Here, $m_{ij}$ are $2\times2$ matrices and $m_{ij}=m_{ji}^{\dagger}$
for the hermicity. Note that $\boldsymbol{Q}_{1}$ and $\bar{\boldsymbol{Q}}_{1}$
are related by the four-fold rotation symmetry $C_{4z}$ along the $z$-axis. By using this, we first classify the density waves with a propagation vector ${\boldsymbol{Q}}_1$ according to the co-little group of ${\boldsymbol{Q}}_1$ which is isomorphic to the point group $D_{\rm{2h}}$. Then, we obtain the density waves with propagation vector $\bar{\boldsymbol{Q}}_1$ by rotating the ${\boldsymbol{Q}}_1$ density waves. Regarding the co-little group of ${\boldsymbol{Q}}_1$, we would like to refer readers to Sec. \ref{subsec:D2h}. 

The bottom line is that we find eight time-reversal symmetry breaking (TRSB) density waves, which we refer to as spin-density waves in the main text, as well as eight time-reversal symmetric (TRS) density waves. These are listed in Sec. \ref{subsec:MofTRSB} and Sec. \ref{subsec:MofTRS}, respectively. 

The RG equations for $\Delta_{\boldsymbol{Q},m}^{({\rm ph})}(y)$
($\boldsymbol{Q}=\boldsymbol{Q}_{1},\bar{\boldsymbol{Q}}_{1}$) can
be obtained by calculating the one-loop corrections to $\Delta_{\boldsymbol{Q},m}^{({\rm ph})}(y)$
from the interactions as we do for the Cooper pairing vertices. Expressing
our interaction as $\hat{V}=\sum_{i=1,\cdots,13}u_{i}\sum_{m}c_{i,a}(\psi^{\dagger}[M_{i,a}]\psi)(\psi^{\dagger}[N_{i,a}]^{\dagger}\psi)$,
the one-loop correction for $\Delta_{\boldsymbol{Q},m}^{({\rm ph})}(y)$
is expressed as $\frac{\partial\Delta_{\boldsymbol{Q},m}^{({\rm ph})}(y)}{\partial y}=[\Gamma_{\boldsymbol{Q}}^{({\rm ph})}(\boldsymbol{u})]_{mm}\Delta_{\boldsymbol{Q},m}^{({\rm ph})}(y)\frac{\partial}{\partial y}\bigg[\sum_{\omega,\boldsymbol{k}}G_{K_{s}}(\boldsymbol{k})G_{K_{t}}(\boldsymbol{k})\bigg]$
with $\boldsymbol{K}_{s}-\boldsymbol{K}_{t}=\pm\boldsymbol{Q}$, $y=\Pi_{\boldsymbol{0}}^{({\rm pp})}(\Omega)/\nu_{0}$. $\Gamma^{\rm{ph}}_{\boldsymbol{Q}}(\boldsymbol{u})$ is obtained by evaluating the first order corrections due to the interactions which yields the expression similar to Eq. \eqref{Gamma_pp}.
Also, we introduced here the nesting parameter $d_{\boldsymbol{Q}}^{({\rm ph})}(y)\equiv\frac{\partial}{\partial y}\bigg[\sum_{\omega,\boldsymbol{k}}G_{K_{s}}(k)G_{K_{t}}(-k)\bigg]$.

$\Gamma_{\boldsymbol{Q},m}^{({\rm ph})}(\boldsymbol{u})$ of the TRSB
density waves for both $\boldsymbol{Q}=\boldsymbol{Q}_{1}$ and $\boldsymbol{Q}=\bar{\boldsymbol{Q}}_{1}$
are given by
\begin{align}
\Gamma_{\boldsymbol{Q},iA_{{\rm u}}}^{({\rm ph})}(\boldsymbol{u})=\Gamma_{\boldsymbol{Q},iB_{{\rm 3g}}}^{({\rm ph})}(\boldsymbol{u})= & 2u_{13}-u_{2}+u_{3}-u_{4}+u_{6},\\
\Gamma_{\boldsymbol{Q},iB_{{\rm 1u}}}^{({\rm ph})}(\boldsymbol{u})=\Gamma_{\boldsymbol{Q},iB_{{\rm 2g}}}^{({\rm ph})}(\boldsymbol{u})= & 2u_{11}-2u_{12}-2u_{13}-u_{2}+u_{3}+u_{4}-u_{6},\\
\{\Gamma_{\boldsymbol{Q},iB_{{\rm 1g}}^{(1)}}^{({\rm ph})}(\boldsymbol{u}),\Gamma_{\boldsymbol{Q},iB_{{\rm 1g}}^{(2)}}^{({\rm ph})}(\boldsymbol{u})\}= & \left(\begin{matrix}2u_{11}-u_{2}-u_{3}-u_{4}-u_{6} & -2u_{5}\\
-2u_{5} & -2u_{12}-u_{2}-u_{3}+u_{4}+u_{6}
\end{matrix}\right),\\
\{\Gamma_{\boldsymbol{Q},iB_{{\rm 2u}}^{(1)}}^{({\rm ph})}(\boldsymbol{u}),\Gamma_{\boldsymbol{Q},iB_{{\rm 2u}}^{(2)}}^{({\rm ph})}(\boldsymbol{u})\}= & \left(\begin{matrix}-2u_{12}-u_{2}-u_{3}+u_{4}+u_{6} & 2u_{5}\\
2u_{5} & 2u_{11}-u_{2}-u_{3}-u_{4}-u_{6}
\end{matrix}\right),\label{eq:Gamma_TRSB}
\end{align}
where $i$ in the subscript of $\Gamma^{({\rm ph})}$ means that these
density waves break the time-reversal symmetry. Here, $B_{{\rm 1g}}^{(a)}$
and $B_{{\rm 3u}}^{(a)}$ with $a=1,2$ denote two distinct matrix
representations behaving like $B_{{\rm 1g}}$ and $B_{{\rm 3u}}$
irreps of $D_{{\rm 2h}}$, repectively. Since they belong to the same
irrep, their linear combinations are possible. The coefficients of
the appropriate linear combinations are given by an eigenvector of
the matrix in Eq. \eqref{eq:Gamma_TRSB}.

As one can see, there are many degeneracies. The degeneracy of $\Gamma_{\boldsymbol{Q},iA_{{\rm u}}}^{({\rm ph})}(\boldsymbol{u})$
and $\Gamma_{\boldsymbol{Q},iB_{{\rm 3g}}}^{({\rm ph})}(\boldsymbol{u})$,
the degeneracy of $\Gamma_{\boldsymbol{Q},iB_{{\rm 1u}}}^{({\rm ph})}(\boldsymbol{u})$
and $\Gamma_{\boldsymbol{Q},iB_{{\rm 2g}}}^{({\rm ph})}(\boldsymbol{u})$,
and the degeneracies of $B_{{\rm 1g}}$ channels and $B_{{\rm 3u}}$
channels. The origin of the degeneracies will be elucidated in Sec.
\ref{subsec:degeneracy}.

Similarly, $\Gamma_{\boldsymbol{Q},m}^{({\rm ph})}(\boldsymbol{u})$
of the TRS density waves are given by

\begin{align}
\Gamma_{\boldsymbol{Q},A_{{\rm u}}}^{({\rm ph})}(\boldsymbol{u})=\Gamma_{\boldsymbol{Q},B_{{\rm 3g}}}^{({\rm ph})}(\boldsymbol{u})= & -2u_{11}+2u_{12}+2u_{13}-u_{2}+u_{3}+u_{4}-u_{6},\\
\Gamma_{\boldsymbol{Q},B_{{\rm 1u}}}^{({\rm ph})}(\boldsymbol{u})=\Gamma_{\boldsymbol{Q},B_{{\rm 2g}}}^{({\rm ph})}(\boldsymbol{u})= & -2u_{13}-u_{2}+u_{3}-u_{4}+u_{6},\\
\{\Gamma_{\boldsymbol{Q},A_{{\rm g}}^{(1)}}^{({\rm ph})}(\boldsymbol{u}),\Gamma_{\boldsymbol{Q},A_{{\rm g}}^{(2)}}^{({\rm ph})}(\boldsymbol{u})\}= & \left(\begin{matrix}2u_{12}-u_{2}-u_{3}+u_{4}+u_{6} & 2u_{5}\\
2u_{5} & -2u_{11}-u_{2}-u_{3}-u_{4}-u_{6}.
\end{matrix}\right),\\
\{\Gamma_{\boldsymbol{Q},B_{{\rm 3u}}^{(1)}}^{({\rm ph})}(\boldsymbol{u}),\Gamma_{\boldsymbol{Q},B_{{\rm 3u}}^{(2)}}^{({\rm ph})}(\boldsymbol{u})\}= & \left(\begin{matrix}-2u_{11}-u_{2}-u_{3}-u_{4}-u_{6} & -2u_{5}\\
-2u_{5} & 2u_{12}-u_{2}-u_{3}+u_{4}+u_{6}
\end{matrix}\right).
\end{align}

Due to this asymptotic behavior of the coupling constants, the vertices
$\Delta_{\boldsymbol{Q},m}^{({\rm ph})}$ diverge like $1/(y_{c}-y)^{\Gamma_{0,m}^{({\rm pp)}}(\boldsymbol{u}^{*})d_{\boldsymbol{Q}}^{({\rm ph})}(y_{c})}$
which leads to $\chi_{\boldsymbol{Q},m}^{({\rm ph})}\sim(y_{c}-y)^{-\alpha}$
with the exponent 
\begin{equation}
\alpha=2\Gamma_{\boldsymbol{Q},m}^{({\rm ph)}}(\boldsymbol{u}^{*})d_{\boldsymbol{Q}}^{({\rm ph})}(y_{c})-1.\label{eq:exponent_DW}
\end{equation}
The channel with the most positive $\Gamma_{\boldsymbol{Q},m}^{({\rm ph)}}(\boldsymbol{u}^{*})d_{\boldsymbol{Q}}^{({\rm ph})}(y_{c})>\frac{1}{2}$
can be the leading instability among the density-waves in the weak-coupling
limit. Here, we would like to emphsize the appearance of the additional
$d_{\boldsymbol{Q}}^{({\rm ph})}(y_{c})$ in Eq. \eqref{eq:exponent_DW}
compared to Eq. \eqref{eq:exponent_SC}.

\subsubsection{$M_{\boldsymbol{Q}_{1},m}^{({\rm ph})}$ and $M_{\bar{\boldsymbol{Q}}_{1},m}^{({\rm ph})}$
of TRSB density waves\label{subsec:MofTRSB}}

Here, the matrix representation $M_{\boldsymbol{Q}_{1},m}^{({\rm ph})}$
of time-reversal symmetry breaking density waves are given, which
are classified according to the symmetries of the group $D_{{\rm 2h}}$.
$M_{\bar{\boldsymbol{Q}}_{1},m}^{({\rm ph})}$ given along with $M_{\boldsymbol{Q}_{1},m}^{({\rm ph})}$
is obtained by applying the four-fold rotation $C_{4z}$ to $M_{\bar{\boldsymbol{Q}}_{1},m}^{({\rm ph})}$.
With $z=\frac{1-i}{\sqrt{2}},$for example, 

\begin{align}
M_{\boldsymbol{Q}_{1}}^{({\rm ph})}(iB_{3g}) & =\left(\begin{array}{c|cccccccc}
 & K_{1,\Uparrow} & K_{1,\Downarrow} & K_{2,\Uparrow} & K_{2,\Downarrow} & K_{3,\Uparrow} & K_{3,\Downarrow} & K_{4,\Uparrow} & K_{4,\Downarrow}\\
\hline K_{1,\Uparrow} &  &  &  &  &  &  &  & z\\
K_{1,\Downarrow} &  &  &  &  &  &  & z^{*}\\
K_{2,\Uparrow} &  &  &  &  &  & z\\
K_{2,\Downarrow} &  &  &  &  & z^{*}\\
K_{3,\Uparrow} &  &  &  & z\\
K_{3,\Downarrow} &  &  & z^{*}\\
K_{4,\Uparrow} &  & z\\
K_{4,\Downarrow} & z^{*}
\end{array}\right),\\
M_{\bar{\boldsymbol{Q}}_{1}}^{({\rm ph})}(i\bar{B}_{{\rm 3g}}) & =\left(\begin{array}{c|cccccccc}
 & K_{1,\Uparrow} & K_{1,\Downarrow} & K_{2,\Uparrow} & K_{2,\Downarrow} & K_{3,\Uparrow} & K_{3,\Downarrow} & K_{4,\Uparrow} & K_{4,\Downarrow}\\
\hline K_{1,\Uparrow} &  &  &  & z^{*}\\
K_{1,\Downarrow} &  &  & z\\
K_{2,\Uparrow} &  & z^{*}\\
K_{2,\Downarrow} & z\\
K_{3,\Uparrow} &  &  &  &  &  &  &  & z^{*}\\
K_{3,\Downarrow} &  &  &  &  &  &  & z\\
K_{4,\Uparrow} &  &  &  &  &  & z^{*}\\
K_{4,\Downarrow} &  &  &  &  & z
\end{array}\right),
\end{align}
where $\ensuremath{\Uparrow}$ and $\Downarrow$ represent the pseudospins
defined by $U(K_{i})$ in Eq. \eqref{eq:pseudospin}. The others are

\begin{align}
M_{\boldsymbol{Q}_{1}}^{({\rm ph})}(iA_{{\rm u}})=\left(\begin{matrix} &  &  &  &  &  &  & -z^{*}\\
 &  &  &  &  &  & z\\
 &  &  &  &  & -z^{*}\\
 &  &  &  & z\\
 &  &  & z^{*}\\
 &  & -z\\
 & z^{*}\\
-z
\end{matrix}\right) & ,\;M_{\bar{\boldsymbol{Q}}_{1}}^{({\rm ph})}(i\bar{A}_{{\rm u}})=\left(\begin{matrix} &  &  & z\\
 &  & -z^{*}\\
 & -z\\
z^{*}\\
 &  &  &  &  &  &  & -z\\
 &  &  &  &  &  & z^{*}\\
 &  &  &  &  & z\\
 &  &  &  & -z^{*}
\end{matrix}\right),\\
M_{\boldsymbol{Q}_{1}}^{({\rm ph})}(iB_{{\rm 1g}}^{(1)})=\left(\begin{matrix} &  &  &  &  &  & z\\
 &  &  &  &  &  &  & -z^{*}\\
 &  &  &  & z^{*}\\
 &  &  &  &  & -z\\
 &  & z\\
 &  &  & -z^{*}\\
z^{*}\\
 & -z
\end{matrix}\right) & ,\;M_{\bar{\boldsymbol{Q}}_{1}}^{({\rm ph})}(i\bar{B}_{{\rm 1g}}^{(1)})=\left(\begin{matrix} &  & -z^{*}\\
 &  &  & z\\
-z\\
 & z^{*}\\
 &  &  &  &  &  & -z^{*}\\
 &  &  &  &  &  &  & z\\
 &  &  &  & -z\\
 &  &  &  &  & z^{*}
\end{matrix}\right),\\
M_{Q_{1}}^{({\rm ph})}(iB_{{\rm 1g}}^{(2)})=\left(\begin{matrix} &  &  &  &  &  & -z^{*}\\
 &  &  &  &  &  &  & z\\
 &  &  &  & -z\\
 &  &  &  &  & z^{*}\\
 &  & -z^{*}\\
 &  &  & z\\
-z\\
 & z^{*}
\end{matrix}\right) & ,\;M_{\bar{Q}_{1}}^{({\rm ph})}(i\bar{B}_{{\rm 1g}}^{(2)})=\left(\begin{matrix} &  & z\\
 &  &  & -z^{*}\\
z^{*}\\
 & -z\\
 &  &  &  &  &  & z\\
 &  &  &  &  &  &  & -z^{*}\\
 &  &  &  & z^{*}\\
 &  &  &  &  & -z
\end{matrix}\right),\\
M_{Q_{1}}^{({\rm ph})}(iB_{{\rm 2g}})=\left(\begin{matrix} &  &  &  &  &  &  & z^{*}\\
 &  &  &  &  &  & z\\
 &  &  &  &  & z^{*}\\
 &  &  &  & z\\
 &  &  & z^{*}\\
 &  & z\\
 & z^{*}\\
z
\end{matrix}\right) & ,\;M_{\bar{Q}_{1}}^{({\rm ph})}(i\bar{B}_{{\rm 2g}})=\left(\begin{matrix} &  &  & z\\
 &  & z^{*}\\
 & z\\
z^{*}\\
 &  &  &  &  &  &  & z\\
 &  &  &  &  &  & z^{*}\\
 &  &  &  &  & z\\
 &  &  &  & z^{*}
\end{matrix}\right),\\
M_{Q_{1}}^{({\rm ph})}(iB_{{\rm 1u}})=\left(\begin{matrix} &  &  &  &  &  &  & -z\\
 &  &  &  &  &  & z^{*}\\
 &  &  &  &  & -z\\
 &  &  &  & z^{*}\\
 &  &  & z\\
 &  & -z^{*}\\
 & z\\
-z^{*}
\end{matrix}\right) & ,\;M_{\bar{Q}_{1}}^{({\rm ph})}(i\bar{B}_{{\rm 1u}})=\left(\begin{matrix} &  &  & -z^{*}\\
 &  & z\\
 & z^{*}\\
-z\\
 &  &  &  &  &  &  & z^{*}\\
 &  &  &  &  &  & -z\\
 &  &  &  &  & -z^{*}\\
 &  &  &  & z
\end{matrix}\right),\\
M_{Q_{1}}^{({\rm ph})}(iB_{{\rm 2u}}^{(1)})=\left(\begin{matrix} &  &  &  &  &  & z\\
 &  &  &  &  &  &  & z^{*}\\
 &  &  &  & -z^{*}\\
 &  &  &  &  & -z\\
 &  & -z\\
 &  &  & -z^{*}\\
z^{*}\\
 & z
\end{matrix}\right) & ,\;M_{\bar{Q}_{1}}^{({\rm ph})}(i\bar{B}_{{\rm 2u}}^{(1)})=\left(\begin{matrix} &  & z^{*}\\
 &  &  & z\\
z\\
 & z^{*}\\
 &  &  &  &  &  & -z^{*}\\
 &  &  &  &  &  &  & -z\\
 &  &  &  & -z\\
 &  &  &  &  & -z^{*}
\end{matrix}\right),\\
M_{Q_{1}}^{({\rm ph})}(iB_{{\rm 2u}}^{(2)})=\left(\begin{matrix} &  &  &  &  &  & -z^{*}\\
 &  &  &  &  &  &  & -z\\
 &  &  &  & z\\
 &  &  &  &  & z^{*}\\
 &  & z^{*}\\
 &  &  & z\\
-z\\
 & -z^{*}
\end{matrix}\right) & ,\;M_{\bar{Q}_{1}}^{({\rm ph})}(i\bar{B}_{{\rm 2u}}^{(2)})=\left(\begin{matrix} &  & -z\\
 &  &  & -z^{*}\\
-z^{*}\\
 & -z\\
 &  &  &  &  &  & z\\
 &  &  &  &  &  &  & z^{*}\\
 &  &  &  & z^{*}\\
 &  &  &  &  & z
\end{matrix}\right).
\end{align}

\subsubsection{$M_{\boldsymbol{Q}_{1},m}^{({\rm ph})}$ and $M_{\bar{\boldsymbol{Q}}_{1},m}^{({\rm ph})}$
of TRS density waves\label{subsec:MofTRS}}

Here, the matrix representation$M_{\boldsymbol{Q}_{1},m}^{({\rm ph})}$
of density waves preserving the time-reversal symmetry are given,
which are classified according to the symmetries of the group $D_{{\rm 2h}}$.
$M_{\bar{\boldsymbol{Q}}_{1},m}^{({\rm ph})}$ given along with $M_{\boldsymbol{Q}_{1},m}^{({\rm ph})}$
is obtained by applying the four-fold rotation $C_{4z}$ to $M_{\bar{\boldsymbol{Q}}_{1},m}^{({\rm ph})}$.
With $z=\frac{1-i}{\sqrt{2}}$,

\begin{align}
M_{\boldsymbol{Q}_{1}}^{({\rm ph})}(B_{{\rm 3g}})=\left(\begin{matrix} &  &  &  &  &  &  & -z\\
 &  &  &  &  &  & z^{*}\\
 &  &  &  &  & z\\
 &  &  &  & -z^{*}\\
 &  &  & -z\\
 &  & z^{*}\\
 & z\\
-z^{*}
\end{matrix}\right) & ,\;M_{\bar{\boldsymbol{Q}}_{1}}^{({\rm ph})}(\bar{B}_{{\rm 3g}})=\left(\begin{matrix} &  &  & z^{*}\\
 &  & -z\\
 & -z^{*}\\
z\\
 &  &  &  &  &  &  & z^{*}\\
 &  &  &  &  &  & -z\\
 &  &  &  &  & -z^{*}\\
 &  &  &  & z
\end{matrix}\right),\\
M_{\boldsymbol{Q}_{1}}^{({\rm ph})}(A_{{\rm u}})=\left(\begin{matrix} &  &  &  &  &  &  & -z^{*}\\
 &  &  &  &  &  & -z\\
 &  &  &  &  & z^{*}\\
 &  &  &  & z\\
 &  &  & z^{*}\\
 &  & z\\
 & -z^{*}\\
-z
\end{matrix}\right) & ,\;M_{\bar{\boldsymbol{Q}}_{1}}^{({\rm ph})}(\bar{A}_{{\rm u}})=\left(\begin{matrix} &  &  & z\\
 &  & z^{*}\\
 & z\\
z^{*}\\
 &  &  &  &  &  &  & -z\\
 &  &  &  &  &  & -z^{*}\\
 &  &  &  &  & -z\\
 &  &  &  & -z^{*}
\end{matrix}\right),\\
M_{\boldsymbol{Q}_{1}}^{({\rm ph})}(B_{{\rm 2g}})=\left(\begin{matrix} &  &  &  &  &  &  & -z^{*}\\
 &  &  &  &  &  & z\\
 &  &  &  &  & z^{*}\\
 &  &  &  & -z\\
 &  &  & -z^{*}\\
 &  & z\\
 & z^{*}\\
-z
\end{matrix}\right) & ,\;M_{\bar{\boldsymbol{Q}}_{1}}^{({\rm ph})}(\bar{B}_{{\rm 2g}})=\left(\begin{matrix} &  &  & -z\\
 &  & z^{*}\\
 & z\\
-z^{*}\\
 &  &  &  &  &  &  & -z\\
 &  &  &  &  &  & z^{*}\\
 &  &  &  &  & z\\
 &  &  &  & -z^{*}
\end{matrix}\right),\\
M_{Q_{1}}^{({\rm ph})}(B_{{\rm 1u}})=\left(\begin{matrix} &  &  &  &  &  &  & z\\
 &  &  &  &  &  & z^{*}\\
 &  &  &  &  & -z\\
 &  &  &  & -z^{*}\\
 &  &  & -z\\
 &  & -z^{*}\\
 & z\\
z^{*}
\end{matrix}\right) & ,\;M_{\bar{Q}_{1}}^{({\rm ph})}(\bar{B}_{{\rm 1u}})=\left(\begin{matrix} &  &  & z^{*}\\
 &  & z\\
 & z^{*}\\
z\\
 &  &  &  &  &  &  & -z^{*}\\
 &  &  &  &  &  & -z\\
 &  &  &  &  & -z^{*}\\
 &  &  &  & -z
\end{matrix}\right),\\
M_{Q_{1}}^{({\rm ph})}(A_{{\rm g}}^{(1)})=\left(\begin{matrix} &  &  &  &  &  & z\\
 &  &  &  &  &  &  & z^{*}\\
 &  &  &  & z^{*}\\
 &  &  &  &  & z\\
 &  & z\\
 &  &  & z^{*}\\
z^{*}\\
 & z
\end{matrix}\right) & ,\;M_{\bar{Q}_{1}}^{({\rm ph})}(\bar{A}_{{\rm g}}^{(1)})=\left(\begin{matrix} &  & z^{*}\\
 &  &  & z\\
z\\
 & z^{*}\\
 &  &  &  &  &  & z^{*}\\
 &  &  &  &  &  &  & z\\
 &  &  &  & z\\
 &  &  &  &  & z^{*}
\end{matrix}\right),\\
M_{Q_{1}}^{({\rm ph})}(A_{{\rm g}}^{(2)})=\left(\begin{matrix} &  &  &  &  &  & z^{*}\\
 &  &  &  &  &  &  & z\\
 &  &  &  & z\\
 &  &  &  &  & z^{*}\\
 &  & z^{*}\\
 &  &  & z\\
z\\
 & z^{*}
\end{matrix}\right) & ,\;M_{\bar{Q}_{1}}^{({\rm ph})}(\bar{A}_{{\rm g}}^{(2)})=\left(\begin{matrix} &  & z\\
 &  &  & z^{*}\\
z^{*}\\
 & z\\
 &  &  &  &  &  & z\\
 &  &  &  &  &  &  & z^{*}\\
 &  &  &  & z^{*}\\
 &  &  &  &  & z
\end{matrix}\right),\\
M_{Q_{1}}^{({\rm ph})}(B_{{\rm 3u}}^{(1)})=\left(\begin{matrix} &  &  &  &  &  & z\\
 &  &  &  &  &  &  & -z^{*}\\
 &  &  &  & -z^{*}\\
 &  &  &  &  & z\\
 &  & -z\\
 &  &  & z^{*}\\
z^{*}\\
 & -z
\end{matrix}\right) & ,\;M_{\bar{Q}_{1}}^{({\rm ph})}(\bar{B}_{{\rm 3u}}^{(1)})=\left(\begin{matrix} &  & z^{*}\\
 &  &  & -z\\
z\\
 & -z^{*}\\
 &  &  &  &  &  & -z^{*}\\
 &  &  &  &  &  &  & z\\
 &  &  &  & -z\\
 &  &  &  &  & z^{*}
\end{matrix}\right),\\
M_{Q_{1}}^{({\rm ph})}(B_{{\rm 3u}}^{(2)})=\left(\begin{matrix} &  &  &  &  &  & -z^{*}\\
 &  &  &  &  &  &  & z\\
 &  &  &  & z\\
 &  &  &  &  & -z^{*}\\
 &  & z^{*}\\
 &  &  & -z\\
-z\\
 & z^{*}
\end{matrix}\right) & ,\;M_{\bar{Q}_{1}}^{({\rm ph})}(\bar{B}_{{\rm 3u}}^{(2)})=\left(\begin{matrix} &  & -z\\
 &  &  & z^{*}\\
-z^{*}\\
 & z\\
 &  &  &  &  &  & z\\
 &  &  &  &  &  &  & -z^{*}\\
 &  &  &  & z^{*}\\
 &  &  &  &  & -z
\end{matrix}\right).
\end{align}

\subsection{Results of parquet RG for several $U$ and $\kappa$}

\begin{figure}
\includegraphics[width=0.99\textwidth]{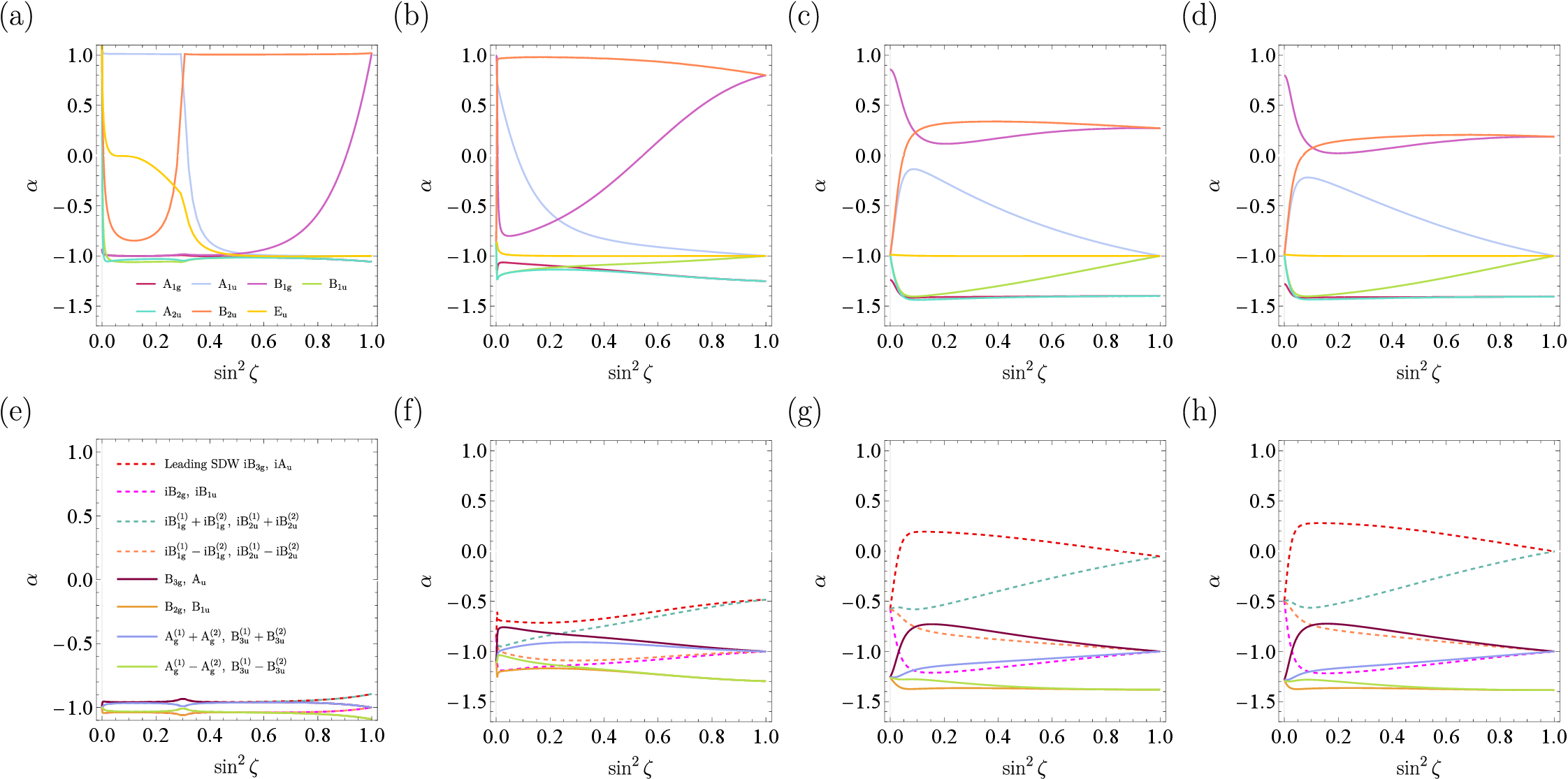}
\caption{\label{fig:S4}The exponent $\alpha$ of susceptibilities of (a-d)
the Cooper pairing channels and (e-h) the density wave channels at
$U=0.02$. For (a) and (e), $\kappa=2$. For (b) and (f), $\kappa=1.1$.
For (c) and (g), $\kappa=1.001$. For (d) and (h), $\kappa=1.0001$.
In (e)-(h), the solid (dashed) lines correspond to the time-reversal
symmetry preserving (breaking) density-wave channels.}
\end{figure}

\begin{figure}
\includegraphics[width=0.99\textwidth]{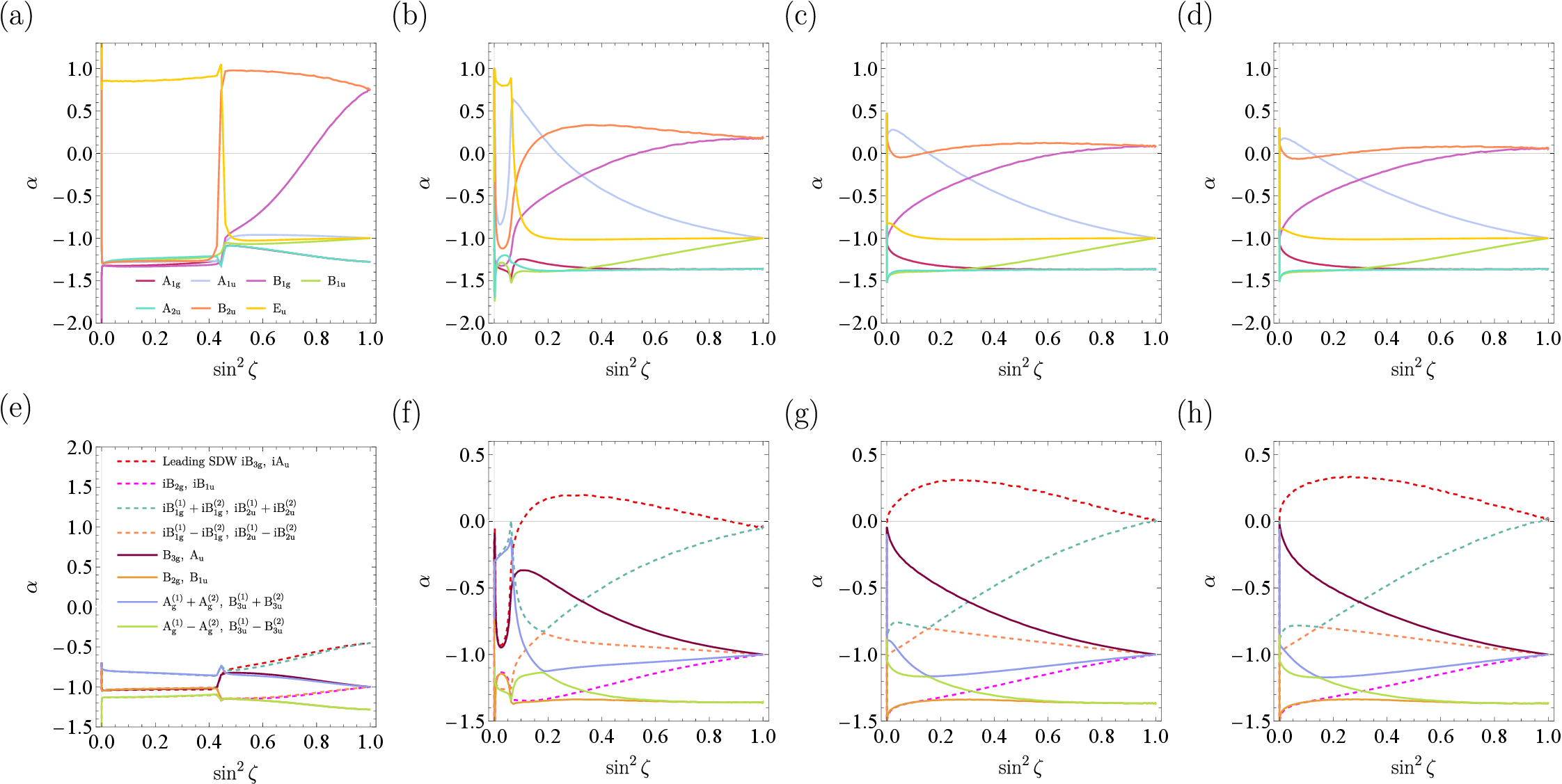}

\caption{\label{fig:S5}The exponent $\alpha$ of susceptibilities of (a-d)
the Cooper pairing channels and (e-h) the density wave channels at
$U=0.2$. For (a) and (e), $\kappa=2$. For (b) and (f), $\kappa=1.1$.
For (c) and (g), $\kappa=1.01$. For (d) and (h), $\kappa=1.001$.
In (e)-(h), the solid (dashed) lines correspond to the time-reversal
symmetry preserving (breaking) density-wave channels.}
\end{figure}

Figures \ref{fig:S4} and \ref{fig:S5} show the exponent $\alpha$
of the Cooper pairing channels and the density wave channels for several
$U$ and $\kappa$ as an evolution of $\sin^{2}\zeta$ in Eq. \eqref{eq:sinzeta}.
For the other channels, we don't see a case with positive $\alpha$.
It is easy to observe that only the Cooper pairing instabilities are
found to have $\Gamma_{0,m}^{({\rm c)}}(\boldsymbol{u}^{*})>0$ or
$\alpha>-1$ when both the bare interaction and the nesting $\kappa$
are small. This is a general consequence of focusing just on the van
Hove singularities which can be understood by looking into RG equations
for the vertices. The equations are generally written as 
\begin{equation}
\frac{\partial\Delta_{\boldsymbol{Q}_{i}}^{(c)}(y)}{\partial y}=\Gamma_{\boldsymbol{Q}_{i}}^{(c)}(\boldsymbol{u})\Delta_{\boldsymbol{Q}_{i}}^{(c)}(y)d_{i}^{(c)}(y)
\end{equation}
with $c={\rm pp},{\rm ph}$. Note the additional $d_{i}^{(c)}(y)$
appears on the right hand side. For the Cooper pairing channels, it
is just $1$. Meanwhile, for all the other channels, $d_{i}^{(c)}(y)$
decreases to $0$ as $y$ increases.

In the weak-coupling limit, the bare interaction is very small. Starting
with small bare couplings results in the very large critical scale
$y_{c}$ at which some coupling constants diverge. Thus, $d_{i}^{({\rm ph})}(y_{c})\ll1$.
As we noted after Eq. \eqref{eq:exponent_DW}, this affects the exponent
$\alpha$ of the susceptibility as well as the vertex. As a result,
this makes other channels less dominant than the Cooper pairing channel
in the weak-coupling limit.

When the nesting is close to being perfect ($\kappa\approx1$), $d_{{\rm 1}}^{(c)}(y)$
for the channels related with $\boldsymbol{Q}=\boldsymbol{Q}_{1}$
are very large at $y=0$. The larger $d_{{\rm 1}}^{(c)}(y)$ let the
coupling constants obtained by solving the RG equations diverge at
smaller $y_{c}$. If $y_{c}\sim O(1)$, this is the case that we see
the postive $\alpha$ of the susceptibility of some density-wave channels.

However, the cases where the bare interaction is strong or $\kappa$
is quite close to $1$ are not the cases in the weak-coupling regime.
Hence, the results obtained with these parameters through the parquet
RG are not secured since the parquet RG just keeps the one-loop corrections
and ignores the corrections to the self-energy which are all essential
in the strong-coupling limit. However, it is often true that what
is obtained through the parquet RG agrees qualitatively with the results
of more sophistcated methods appropriate for the intermediate or strong
coupling regime \citep{Meng2015}. 

\section{Further analysis of the density wave orders}
\label{sec:density-wave}
Unlike the Cooper pairings with zero total momentum, the density wave
orders involve finite momentum transferred by a particle-hole pair.
As we mentioned, this complicates the symmetry classification of these
channels, which requires a more or less lengthy delineation. This
this section is prepared for this delineation which is necessary to
understand the degeneracies between various density-wave channels
in our RG results.

\subsection{Derivation of Eq. \eqref{eq:V_DW}}
\label{subsec:derivation}

We commence by deriving Eq. \eqref{eq:V_DW} since this will help
to understand the most common type of degeneracies in our RG results. From the mean-field point of view, let us suppose that there is a
potential oscillating like a cosine function over the crystal: 
\begin{equation}
\Delta(\boldsymbol{r})=2{\cal M}(\boldsymbol{r})\cos\boldsymbol{Q}\cdot\boldsymbol{r}.
\end{equation}
${\cal M}(\boldsymbol{r})$ is generally a hermitian matrix-valued function periodic in the original lattice.

\begin{figure}
\includegraphics[width=0.3\columnwidth]{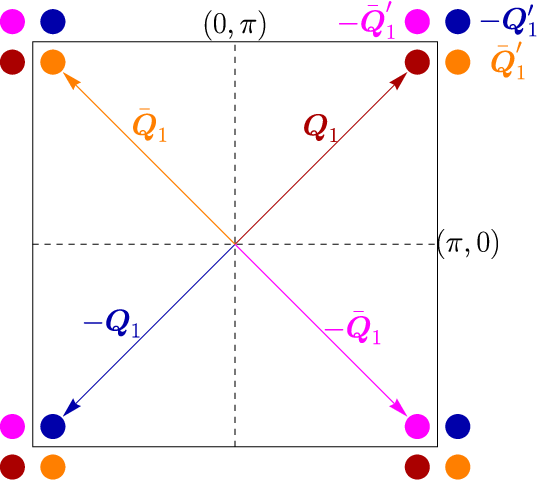}
\caption{\label{fig:S6}Sets of equivalent wave vectors. The colored dots represent
the wave vectors. Same color means the equivalece.}
\end{figure}

By using the creation and annihilation operators of electrons denoted
by $\hat{\Psi}^{\dagger}(\boldsymbol{r})$ and $\hat{\Psi}(\boldsymbol{r})$,
respectively, the term added to the total Hamiltonian due to $\Delta(\boldsymbol{r})$
is given by
\begin{equation}
\hat{V}_{{\rm cos-DW}}=\int_{\boldsymbol{r}}\hat{\Psi}^{\dagger}(\boldsymbol{r})\Delta(\boldsymbol{r})\hat{\Psi}(\boldsymbol{r}).
\end{equation}
To derive Eq. \eqref{eq:V_DW}, we need to use the expression of $\hat{\Psi}(\boldsymbol{r})$
in terms of the field operators in the Wannier basis as well as the
Bloch basis which are written as 
\begin{equation}
\hat{\Psi}(\boldsymbol{r})=\sum_{\boldsymbol{R}}\sum_{a=A,B}\varphi(\boldsymbol{r}-\boldsymbol{R}-\boldsymbol{b}_{a})\hat{C}_{a}(\boldsymbol{R})=\sum_{\boldsymbol{k}\in{\rm FBZ}}\sum_{a=A,B}\bigg[\sum_{\boldsymbol{R}}\varphi(\boldsymbol{r}-\boldsymbol{R}-\boldsymbol{b}_{a})e^{-i\boldsymbol{k}\cdot(\boldsymbol{R}+\boldsymbol{b}_{a})}\bigg]\hat{C}_{a}(\boldsymbol{k}),
\end{equation}
with the lattice vectors $\boldsymbol{R}$. $\boldsymbol{b}_{a}$
($a=A,B$) denote the position of each sublattice with respect to
the origin inside the primitive unit cell at the lattice site $\boldsymbol{R}$.
For simplicity, we considered a single $s$-wave like orbital $\varphi$
at each site. The expression of $\hat{V}_{{\rm DW}}$ in the Wannier
basis is written as
\begin{align}
\hat{V}_{{\rm cos-DW}}= & \sum_{\boldsymbol{R},a}\sum_{\boldsymbol{R}',a'}\hat{C}_{a}^{\dagger}(\boldsymbol{R})\hat{C}_{a'}(\boldsymbol{R}')\int_{\boldsymbol{r}}\varphi^{*}(\boldsymbol{r}-\boldsymbol{R}-\boldsymbol{b}_{a})\Delta(\boldsymbol{r})\varphi(\boldsymbol{r}-\boldsymbol{R}'-\boldsymbol{b}_{a'})\nonumber \\
= & \sum_{\xi=\pm}\sum_{\boldsymbol{R},a}\sum_{\boldsymbol{R}',a'}\hat{C}_{a}^{\dagger}(\boldsymbol{R})\hat{C}_{a'}(\boldsymbol{R}'+\boldsymbol{R})e^{i\xi\boldsymbol{Q}\cdot(\boldsymbol{R}+\boldsymbol{b}_{a})}\int_{\boldsymbol{r}}\varphi^{*}(\boldsymbol{r}){\cal M}(\boldsymbol{r})\varphi(\boldsymbol{r}-\boldsymbol{R}'-\boldsymbol{b}_{a'}+\boldsymbol{b}_{a}).
\end{align}
For the further simplication, we assume that $M_{aa'}(\boldsymbol{R})=\int_{\boldsymbol{r}}\varphi^{*}(\boldsymbol{r}){\cal M}(\boldsymbol{r})\varphi(\boldsymbol{r}-\boldsymbol{R}-\boldsymbol{b}_{a'}+\boldsymbol{b}_{a})$
is finite only when $\boldsymbol{R}'=\boldsymbol{0}$ and $\boldsymbol{b}_{a'}=\boldsymbol{b}_{a}$,
which means that the density-wave gives rise to an oscillating local
potential.
\begin{align}
\hat{V}_{{\rm cos-DW}}= & \sum_{\boldsymbol{R},a}\left(\begin{matrix}\hat{C}_{A}^{\dagger}(\boldsymbol{R}) & \hat{C}_{B}^{\dagger}(\boldsymbol{R})\end{matrix}\right)M(\boldsymbol{R})\left(\begin{matrix}\hat{C}_{A}(\boldsymbol{R})\\
\hat{C}_{B}(\boldsymbol{R})
\end{matrix}\right),\\
M(\boldsymbol{R})= & \left(\begin{matrix}M_{AA}\cos\boldsymbol{Q}\cdot(\boldsymbol{R}+\boldsymbol{b}_{A})\\
 & M_{BB}\cos\boldsymbol{Q}\cdot(\boldsymbol{R}+\boldsymbol{b}_{B})
\end{matrix}\right).
\end{align}
Transforming into the expression using the Bloch basis is straightforward. 

\begin{align}
\hat{V}_{{\rm cos-DW}}= & \sum_{\boldsymbol{k}\in{\rm FBZ}}\hat{C}_{\boldsymbol{k}}^{\dagger}M\hat{C}_{\boldsymbol{k}+\boldsymbol{Q}}+\sum_{\boldsymbol{k}\in{\rm FBZ}}\hat{C}_{\boldsymbol{k}+\boldsymbol{Q}}^{\dagger}M\hat{C}_{\boldsymbol{k}}\\
= & \sum_{\boldsymbol{k}\in{\rm FBZ}}\left(\begin{matrix}\hat{C}_{\boldsymbol{k}}^{\dagger} & \hat{C}_{\boldsymbol{k}+\boldsymbol{Q}}^{\dagger}\end{matrix}\right)\left(\begin{matrix} & M\\
M
\end{matrix}\right)\left(\begin{matrix}\hat{C}_{\boldsymbol{k}}\\
\hat{C}_{\boldsymbol{k}+\boldsymbol{Q}}
\end{matrix}\right)\label{eq:cos-DW}
\end{align}
where $\hat{C}_{\boldsymbol{k}}^{\dagger}=\left(\begin{matrix}\hat{C}_{A}^{\dagger}(\boldsymbol{k}) & \hat{C}_{B}^{\dagger}(\boldsymbol{k})\end{matrix}\right)$
and $M={\rm diag}(M_{AA},M_{BB})$. Here, we used $\hat{C}_{a}(\boldsymbol{k}+\boldsymbol{G})=e^{i\boldsymbol{G}\cdot\boldsymbol{b}_{a}}\hat{C}_{a}(\boldsymbol{k})$
in deriving the first line. Eq. \eqref{eq:V_DW} is derived by projecting
$\hat{V}_{{\rm cos-DW}}$ onto the low-energy states at the patches.
Note that if we take $\boldsymbol{Q}'=\boldsymbol{Q}+\boldsymbol{G}$
with a reciprocal lattice vector $\boldsymbol{G}$ as the wave vector
of the density wave, we get
\begin{align}
\hat{V}_{{\rm cos-DW}}= & \sum_{\boldsymbol{k}\in{\rm FBZ}}\left(\begin{matrix}\hat{C}_{\boldsymbol{k}}^{\dagger} & \hat{C}_{\boldsymbol{k}+\boldsymbol{Q}'}^{\dagger}\end{matrix}\right)\left(\begin{matrix} & MV_{\boldsymbol{G}}\\
V_{\boldsymbol{G}}^{\dagger}M
\end{matrix}\right)\left(\begin{matrix}\hat{C}_{\boldsymbol{k}}\\
\hat{C}_{\boldsymbol{k}+\boldsymbol{Q}'}
\end{matrix}\right)
\end{align}
with $V_{\boldsymbol{G}}={\rm diag}(e^{-i\boldsymbol{G}\cdot\boldsymbol{b}_{A}},e^{-i\boldsymbol{G}\cdot\boldsymbol{b}_{B}})\otimes s_{0}$
where $s_{0}$ is the identity matrix in the spin space. As $M\cos(\boldsymbol{Q}\cdot\boldsymbol{r})$
and $(MV_{\boldsymbol{G}}e^{i\boldsymbol{Q}'\cdot\boldsymbol{r}}+V_{\boldsymbol{G}}^{\dagger}Me^{-i\boldsymbol{Q}'\cdot\boldsymbol{r}})/2$
are equivalent on the lattice, both expressions of $\hat{V}_{{\rm cos-DW}}$
are also equivalent. Figure \ref{fig:S6} shows the four sets of equivalent
wave vector $\boldsymbol{Q}$ of density waves.

We would like to note here that the simple-looking of Eq. \eqref{eq:cos-DW} is somewhat deceiving. It is a shorthand expression of the following long string (possibly infinitely long when $\boldsymbol{Q}$ is truly incommensurate):
\begin{align}
\hat{V}_{{\rm cos-DW}}=\sum_{\boldsymbol{k}\in{\rm RBZ}}\{\hat{C}_{\boldsymbol{k}}^{\dagger}M\hat{C}_{\boldsymbol{k}+\boldsymbol{Q}}+\hat{C}_{\boldsymbol{k}+\boldsymbol{Q}}^{\dagger}M\hat{C}_{\boldsymbol{k}+2\boldsymbol{Q}}+\hat{C}_{\boldsymbol{k}+2\boldsymbol{Q}}^{\dagger}M\hat{C}_{\boldsymbol{k}+3\boldsymbol{Q}}+\cdots\}
\end{align}
for general $\boldsymbol{Q}$. Here, RBZ means the reduced Brillouin zone due to the density wave with propagation vector $\boldsymbol{Q}$.

If $\Delta(\boldsymbol{r})={\cal M}(\boldsymbol{r})\sin(\boldsymbol{Q}\cdot\boldsymbol{r})$,
the Bloch basis representation of it is given by 
\begin{align}
\hat{V}_{{\rm sin-DW}}= & \sum_{\boldsymbol{k}\in{\rm FBZ}}\left(\begin{matrix}\hat{C}_{\boldsymbol{k}}^{\dagger} & \hat{C}_{\boldsymbol{k}+\boldsymbol{Q}}^{\dagger}\end{matrix}\right)\left(\begin{matrix} & -iM\\
iM
\end{matrix}\right)\left(\begin{matrix}\hat{C}_{\boldsymbol{k}}\\
\hat{C}_{\boldsymbol{k}+\boldsymbol{Q}}
\end{matrix}\right).
\end{align}

Note that $\hat{V}_{{\rm cos-DW}}$ is transformed into $\hat{V}_{{\rm sin-DW}}$
through a gauge transformation 
\begin{equation}
(\hat{C}_{\boldsymbol{k}},\hat{C}_{\boldsymbol{k}+\boldsymbol{Q}})\rightarrow(e^{i\Theta(\boldsymbol{k})}\hat{C}_{\boldsymbol{k}},e^{i\Theta(\boldsymbol{k}+\boldsymbol{Q})}\hat{C}_{\boldsymbol{k}+\boldsymbol{Q}})
\end{equation}
 where the function $\Theta(\boldsymbol{k})$ satisfies 
\begin{align}
\Theta(\boldsymbol{k}) & =-\Theta(-\boldsymbol{k}),\\
\Theta(\boldsymbol{k}+\boldsymbol{Q}) & =\Theta(\boldsymbol{k})+\pi/2.
\end{align}
An example is $\Theta(\boldsymbol{k})=\frac{1}{4}\boldsymbol{R}\cdot\boldsymbol{k}$
with $\boldsymbol{R}\cdot\boldsymbol{Q}=2\pi$. As presented in Sec.
\ref{subsec:multiple_degeneracy_of_DW}, this symmetry is related
with a translation which is broken in the presence of
the spin-density wave with general $\boldsymbol{Q}$.

\subsection{\label{subsec:D2h}The co-little group of ${\boldsymbol{Q}}_{1}$ and $-{\boldsymbol{Q}}_{1}$ in the non-symmorphic space group $P4/nmm$}

In the space group $P4/nmm$, the inversion transformation and two-fold
rotation transformations are accompanied by half-lattice translations
when we choose the rotation axis of a four-fold rotation as the origin
of the referential frame. Using the notation $\{R|\boldsymbol{t}\}\boldsymbol{r}=R\boldsymbol{r}+\boldsymbol{t}$
with a proper or improper rotation $R$ and a translation $t$, an
inversion transformation and two rotations around $(110)$ direction
and $(1\bar{1}0)$ direction in the space group $P4/nmm$ can be expressed
as $\{{\cal I}|\bar{\boldsymbol{\tau}}\}$, $\{2_{110}|\bar{\boldsymbol{\tau}}\}$,
and $\{2_{1\bar{1}0}|\bar{\boldsymbol{\tau}}\}$ with $\bar{\boldsymbol{\tau}}=(\frac{1}{2},-\frac{1}{2},0)$
is the accompanied half-lattice translation. Let $G$ denote the product
of the group generated by $\{\{{\cal I}|\bar{\boldsymbol{\tau}}\},\{2_{110}|\bar{\boldsymbol{\tau}}\},\{2_{1\bar{1}0}|\bar{\boldsymbol{\tau}}\}\}$
and the group $T$ generated by $\{\{E|100\},\{E|010\}\}$. The coset
$G/T\sim\bar{D}$ is a quotient group, where $\bar{D}$ is a set of eight
elements:
\begin{equation}
\bar{D}=\{\{E|\boldsymbol{0}\},\{{\cal I}|\bar{\boldsymbol{\tau}}\},\{2_{110}|\bar{\boldsymbol{\tau}}\},\{2_{1\bar{1}0}|\bar{\boldsymbol{\tau}}\},\{2_{001}|\boldsymbol{0}\},\{m_{110}|\boldsymbol{0}\},\{m_{1\bar{1}0}|\boldsymbol{0}\},\{m_{001}|\bar{\boldsymbol{\tau}}\}\},
\end{equation}
in which all elements involve lattice translation satisfying $\boldsymbol{Q}_{1}\cdot\bar{\boldsymbol{\tau}}=0$. It is important to note $\bar{D}$ is isomorphic to the crystallographic point group $D_{{\rm 2h}}$. Consequently, we can regard the quotient group $\bar{D}$ as the normal point group $D_{{\rm 2h}}$ and we can use the familiar irreducible representations of $D_{{\rm 2h}}$ to classify the density waves with a propagation vector $\boldsymbol{Q}_{1}=(k,k)$.


The co-little group of the momentum $\bar{\boldsymbol{Q}}_1$ and $-\bar{\boldsymbol{Q}}_1$ can be expressed as $D\equiv C_{4,z}\bar{D}C_{4,z}^{-1}$ with $C_{4,z}$ the 90$^\circ$ rotation about the $z$-axis. As a result, every density wave functions with a propagation vector ${\boldsymbol{Q}}_1$ has a degenerate partner obtained by rotating it 90$^\circ$ about the $z$-axis.



\subsection{Multiple degeneracy of density
waves}
\label{subsec:multiple_degeneracy_of_DW}
\begin{figure}
\includegraphics[width=0.5\linewidth]{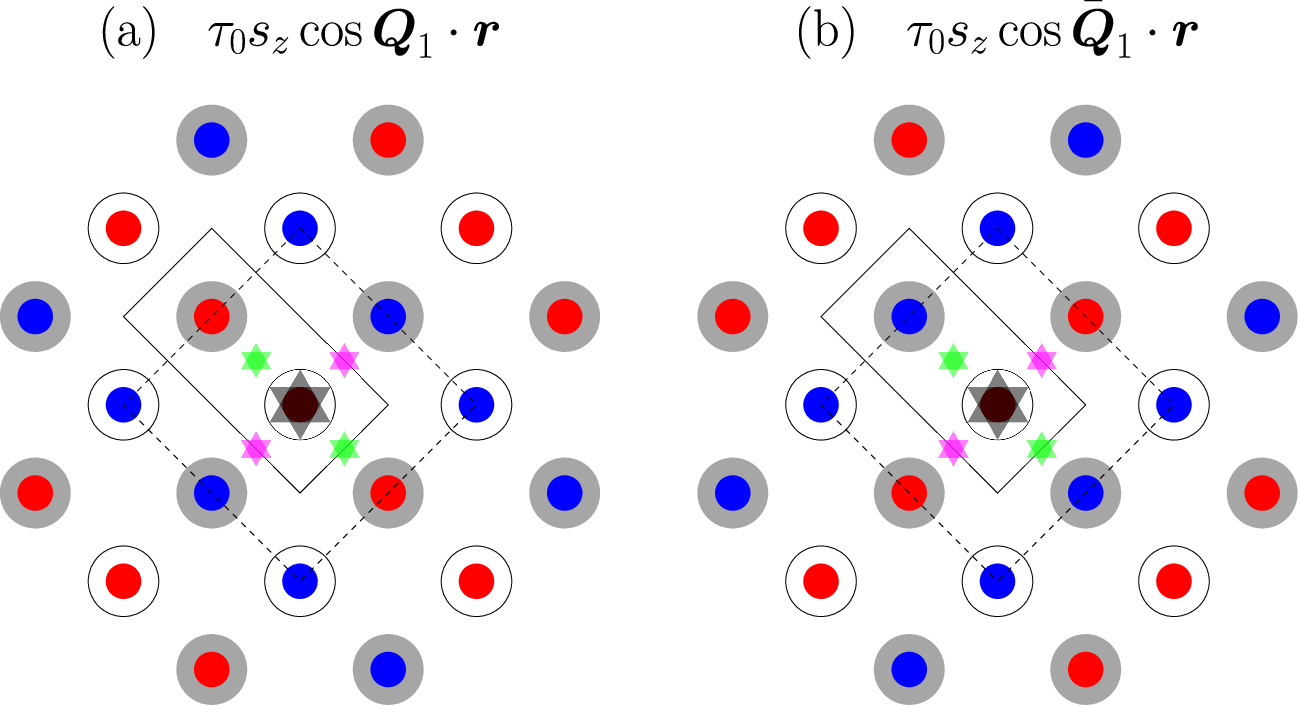}
\caption{\label{fig:S7}Top view of the lattice and the local magnetic moments
of a spin density wave. (a) The configuration of the local magnetic
moment at the sublattices due to $\Delta_{{\rm DW}}(\boldsymbol{r})=\tau_{0}s_{z}\cos(\boldsymbol{Q}_{1}\cdot\boldsymbol{r})$
with $\boldsymbol{Q}_{1}=(\pi,\pi)$. Red dots and blue dots represent the magnetic moment pointing the $+z$ and $-z$ direction, respectively. (b) The configuration of the local magnetic moment at the sublattices due to $\bar{\Delta}_{{\rm DW}}(\boldsymbol{r})=\tau_{0}s_{z}\cos(\bar{\boldsymbol{Q}}_{1}\cdot\boldsymbol{r})$
with $\bar{\boldsymbol{Q}}_{1}=(-\pi,\pi)$. In (a) and (b), the tilted
black box represents the primitive unit cell of the original lattice.
Black star marks the origin of the reference frame viewed from the
top. The magenta and green stars around the origin are representative
inversion centers of the original lattice. The $\sqrt{2}\times\sqrt{2}$
magnetic unit cell in the presence of a density wave with $\boldsymbol{Q}=(\pm\pi,\pi)$
is depicted by a black dashed diamond.}
\end{figure}


In this section, we explain how the classification using the co-little groups $\bar{D}$ and $D$ of the propagation vectors leads to the four-fold degenerate density waves for general ${\boldsymbol{Q}}_1=(k,k)$ with $0<k<\pi$. In addition, we also provide how the so-called \emph{improper} irreducible representation \citep{Szabo2023} can occur at $k=\pi$, in which an even parity density wave and an odd-parity density wave form the basis together. This also has been investigated in Refs. \citep{Serbyn2013,Cvetkovic2013,Venderbos2016,Szabo2023}.

To make the point easy, let us use two spin density waves of out-of-plane local magnetic moments, which are represented by $\tau_{0}s_{z}\cos(\boldsymbol{Q}\cdot\boldsymbol{r})$ and $\tau_{0}s_{z}\cos(\bar{\boldsymbol{Q}}\cdot\boldsymbol{r})$. Figures \ref{fig:S7}(a) and \ref{fig:S7}(b) illustrate the spin configuration of these spin density waves, respectively, viewed from the top of the lattice. Two directions of the out-of-plane magnetic moments are represented by the red and blue dots. The empty (filled) disk represents a sublattice A (B) located at $z=-\frac{1}{2}$ ($z=\frac{1}{2}$).
The tilted box depicts a primitive unit cell containing a sublattice A at $(0,0,-\frac{1}{2})$ and a sublattice B at $(-\frac{1}{2},\frac{1}{2},\frac{1}{2})$.
The projection of the origin of the reference frame is marked by a black star, which serves as the axis for rotations about the $z$-axis. The magenta stars and green stars mark the center of inversion around the origin. In the present reference frame, the inversion transformation with respect to a green star and a magenta star can be written as $\{{\cal I}|\tau\}$ and $\{{\cal I}|\bar{\tau}\}$, respectively where $\tau=(\frac{1}{2},\frac{1}{2},0)$,
$\bar{\tau}=(-\frac{1}{2},\frac{1}{2},0)$, and ${\cal I}$ means
the inversion in the three dimensional real space $\mathbb{R}^{3}$.
Black diamond describes a magnetic unit-cell when there is a density
wave with $\boldsymbol{Q}=(\pi,\pi)$. As one can see, the two inversion centers become inequivalent when the density waves are developed which break the translation symmetries  $\{E|100\}$ and $\{E|010\}$.

The presence of two distinct inversion center in a magnetic unit cell leads to the \emph{improper} irreducible representation. Note that the spin density wave $\tau_{0}s_{z}\cos(\boldsymbol{Q}\cdot\boldsymbol{r})$ in Fig. \ref{fig:S7}(a) is even under the inversion ($\{{\cal I}|\bar{\tau}\}$) around the green stars, while $\tau_{0}s_{z}\cos(\bar{\boldsymbol{Q}}\cdot\boldsymbol{r})\}$ is odd under the same inversion. As these are related by a four-fold rotation, an irreducible representation should contain these two spin-density waves together as the basis of the representation. Therefore, we can say that the irreducible representation represented by $\{\tau_{0}s_{z}\cos(\boldsymbol{Q}\cdot\boldsymbol{r}),\tau_{0}s_{z}\cos(\bar{\boldsymbol{Q}}\cdot\boldsymbol{r})$ mixes the parity, which is why this irreducible representation is described as \emph{improper} in Ref. \cite{Szabo2023}. The goes for any density waves with a propagation vector $(\pm \pi,\pi)$, and thus we can generally expect two dimensional irreducible representations of density waves with propagation vectors $(\pm \pi,\pi)$.

For general $\boldsymbol{Q}=(k,k)$ with $0<k<\pi$, however, more higher-dimensional irreducible representations are derived. In this case, propagation vectors with $0<k<\pi$, we can see that $\cos$-like oscillations and $\sin$-like oscillations are degenerate because of the translation symmetries ${E|100}$ and ${E|010}$ of the original lattice. To be specific,
\begin{equation}
\tau_{0}s_{z}\cos(\boldsymbol{Q}_{1}\cdot(\boldsymbol{r}-(1,0,0))=\tau_{0}s_{z}\cos(\boldsymbol{Q}_{1}\cdot\boldsymbol{r}-k)=\tau_{0}s_{z}\{\cos k\cos(\boldsymbol{Q}_{1}\cdot\boldsymbol{r})+\sin k\sin(\boldsymbol{Q}_{1}\cdot\boldsymbol{r})\}
\end{equation}
implies that $\tau_{0}s_{z}\cos(\boldsymbol{Q}_{1}\cdot\boldsymbol{r})$,
$\tau_{0}s_{z}\sin(\boldsymbol{Q}_{1}\cdot\boldsymbol{r})$, $\tau_{0}s_{z}\cos(\bar{\boldsymbol{Q}}_{1}\cdot\boldsymbol{r})$
and $\tau_{0}s_{z}\sin(\bar{\boldsymbol{Q}}_{1}\cdot\boldsymbol{r})$
should be degenerate. Therefore, four dimensional irreducible representations are derived for general ${\boldsymbol{Q}}=(\pm k,k)$.

It should be noted that each of $\tau_{0}s_{z}\cos(\boldsymbol{Q}_{1}\cdot\boldsymbol{r})$ and
$\tau_{0}s_{z}\sin(\boldsymbol{Q}_{1}\cdot\boldsymbol{r})$ can be a basis for distinct irreducible representations of $\bar{D} \sim D_{\rm{2h}}$, and thus a pair of irreducible representations found in the $D_{\rm{2h}}$ classification should be degenerate when $0<k<\pi$ as shown in Fig. \ref{fig:S4} and Fig. \ref{fig:S5}. 

\subsection{Density waves in real space}

\begin{figure}
\includegraphics[width=1\linewidth]{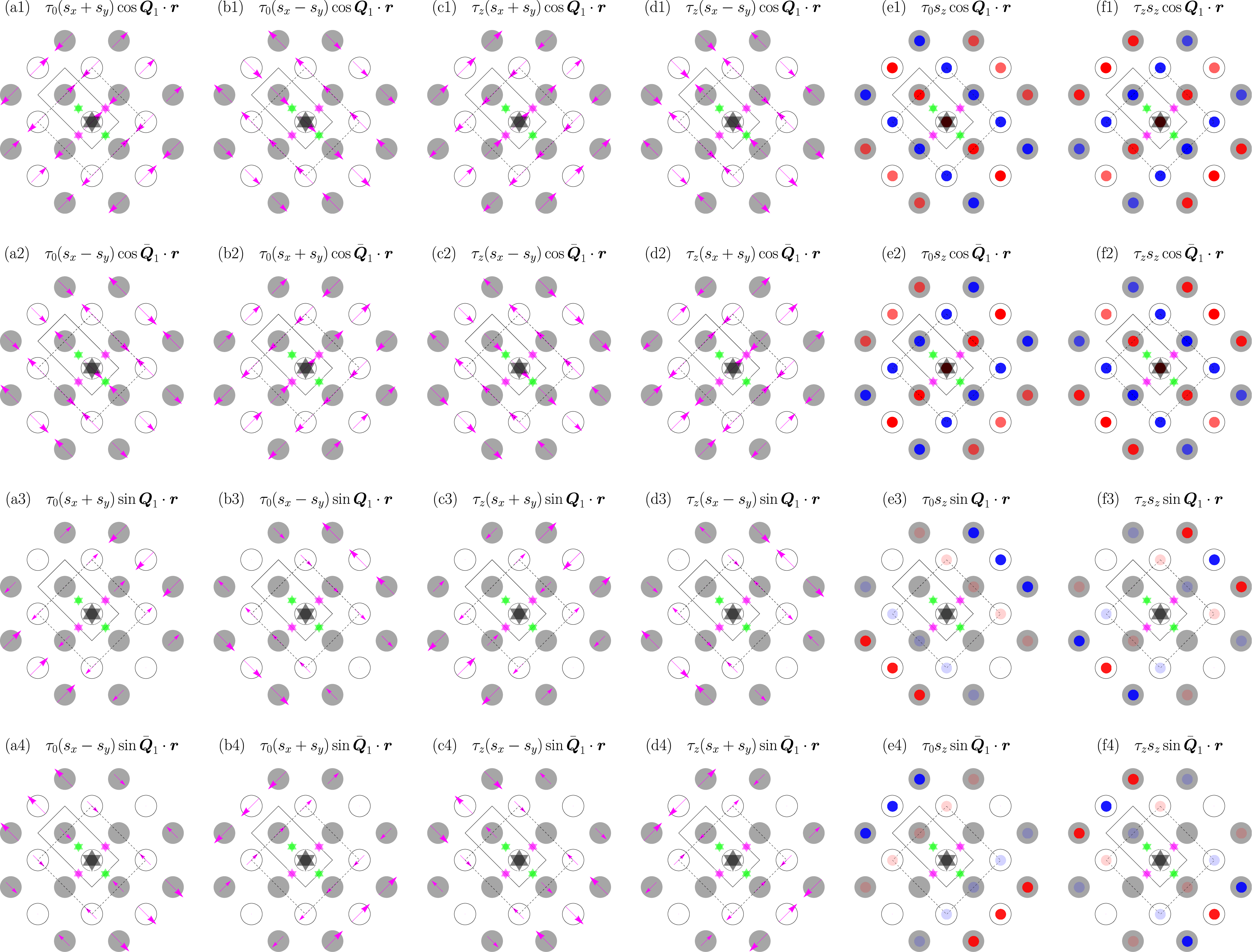}
\caption{\label{fig:S8} The local magnetic moments at the sites of the sublattices
due to the time-reversal symmetry breaking density waves with $\boldsymbol{Q}=0.8(\pm\pi,\pi)$.
The pink arrows represent the in-plane magnetic moments. Red dots
and blue dots mean that the magnetic moment is polarized along the
$+z$ and $-z$ direction, respectively. The magnitude of the magnetic
moments are represented by the length of the arrows or the opacitiy
of the dots. Four density waves in the same column are degenerate.}
\end{figure}

The time-reversal symmetry breaking or preserving density waves in Sec. \ref{subsec:DW} can be associated with spin density waves (SDW) and charge density waves (CDW) in the real space by projecting them onto the low-energy subspace defined by the electronic states at the VHSs near the Fermi level. We focus on the spin-density waves here as CDWs never appear with positive exponents $\alpha$ in our RG analysis. The configuration of local magnetic moments at the Ce sites due to various spin density waves are shown in Figure \ref{fig:S8} where we use $\boldsymbol{Q}=0.8(\pm\pi,\pi)$ for illustration. Four density waves in the same column in Fig. \ref{fig:S8} are degenerate for general $\boldsymbol{Q}=(\pm k,k)$ as we explained in the previous section.

\begin{figure}
\includegraphics[width=1\linewidth]{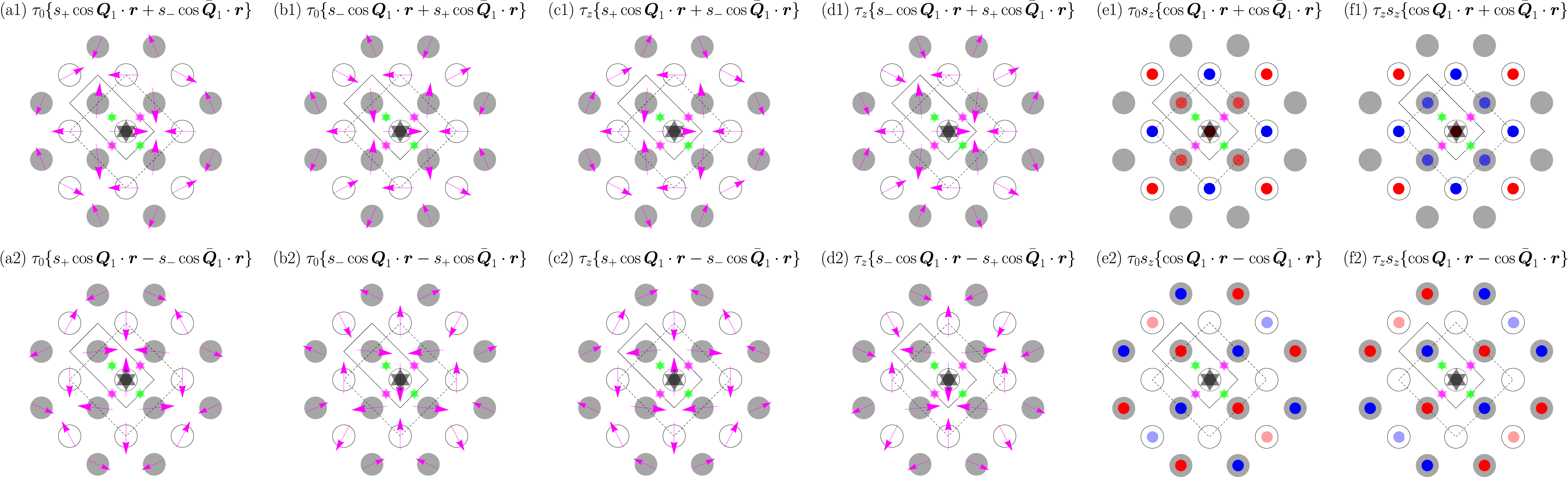}
\caption{\label{fig:S9} The local magnetic moments of the spin density waves defined as the linear combination of SDWs displayed in the first two rows in Fig. \ref{fig:S8}. Here, $s_\pm=s_x\pm s_y$. In (c1), (c2), (d1), and (d2), the circles without red or blue dot represent sublattices where the charge is evacuated.}
\end{figure}

Figure \ref{fig:S9} shows the configuration of local magnetic moments obtained by the symmetric/antisymmetric linear combinations of two SDWs with $\cos$-like oscillation displayed in the first two rows of Fig. \ref{fig:S8}. In the first and the fourth coulums, the spin configuration exhibts hedge-hog like spin texture, while the SDWs in the second and the third columns have vortex cores. The last two columns correspond to the spin-charge density waves with out-of-plane local magnetic moments.

Table \ref{tab:3} summarizes the results of the projection of the density waves with wave vector $\boldsymbol{Q}_{1}$. Note that the SDWs with form factor $M=\tau_z (s_x-s_y)$ are completely suppressed in the low-energy subspace even though they have the in-plane spin texture similar to that of the SDWs with form factor $M=\tau_0 (s_x+s_y)$. This is because the spin-rotation symmetry is broken in our system.
$M_{\boldsymbol{Q}_{1},m}^{({\rm ph})}$ in the third column appearing multiplied by $\sin\zeta$ ($\cos\zeta$) are caused by intra-layer (inter-layer) scatterings through the density wave in the first column. Thus, when there is no effect of the interlayer hopping ($\cos\zeta=0$), such density wave channels are completely suppressed. For example, the exponent $\alpha$ of $M_{\boldsymbol{Q}_{1}}^{({\rm ph})}(iB_{{\rm 2g}})$ is always $-1$ in Fig. \ref{fig:S4} and Fig. \ref{fig:S5}, which
means that the corresponding SDW channels for it does not receive any one-loop correction
through the RG iteration if we start with the on-site Hubbard interaction.

It is worthwhile to note that $\tau_{0}(s_{x}+s_{y})\cos(\boldsymbol{Q}_{1}\cdot\boldsymbol{r})$
is reduced into $\sqrt{2}M_{\boldsymbol{Q}_{1}}^{({\rm ph})}(iB_{{\rm 3g}})$
while $\tau_{z}s_{z}\cos(\boldsymbol{Q}_{1}\cdot\boldsymbol{r})$
is reduced into $\frac{\sin\zeta}{\sqrt{2}}M_{\boldsymbol{Q}_{1}}^{({\rm ph})}(iA_{{\rm u}})$ in the limit ${\boldsymbol{Q}}_1\rightarrow(\pi,\pi)$. 
The norm of the result of the projection is different by a factor
of $\frac{1}{2}\sin\zeta$. Considering how the transition temperature
is calculated in the self-consistent mean-field theory, this difference
in the overall factor will make the antiferromagnetism with $\tau_{0}(s_{x}+s_{y})\cos(\boldsymbol{Q}_{1}\cdot\boldsymbol{r})$
favored over the antiferromagnetism due to $\tau_{z}s_{z}\cos(\boldsymbol{Q}_{1}\cdot\boldsymbol{r})$.

\begin{table}
\begin{tabular}{|c|c|c|}
\hline 
$M$ & cos/sin & Projection to the low-energy states\tabularnewline
\hline 
\hline 
\multirow{2}{*}{$\tau_{0}(s_{x}+s_{y})$} & (a1) $\cos(\boldsymbol{Q}_{1}\cdot\boldsymbol{r})$ & $\sqrt{2}M_{\boldsymbol{Q}_{1}}^{({\rm ph})}(iB_{{\rm 3g}})$\tabularnewline
\cline{2-3} \cline{3-3} 
 & (a3) $\sin(\boldsymbol{Q}_{1}\cdot\boldsymbol{r})$ & $\sqrt{2}M_{\boldsymbol{Q}_{1}}^{({\rm ph})}(iA_{{\rm u}})$\tabularnewline
\hline 
\multirow{2}{*}{$\tau_{0}(s_{x}-s_{y})$} & (b1) $\cos(\boldsymbol{Q}_{1}\cdot\boldsymbol{r})$ & $\sqrt{2}\cos\zeta\cdot M_{\boldsymbol{Q}_{1}}^{({\rm ph})}(iB_{{\rm 2g}})$\tabularnewline
\cline{2-3} \cline{3-3} 
 & (b3) $\sin(\boldsymbol{Q}_{1}\cdot\boldsymbol{r})$ & $-\sqrt{2}\cos\zeta\cdot M_{\boldsymbol{Q}_{1}}^{({\rm ph})}(iB_{{\rm 1u}})$\tabularnewline
\hline 
\multirow{2}{*}{$\tau_{z}(s_{x}+s_{y})$} & (c1) $\cos(\boldsymbol{Q}_{1}\cdot\boldsymbol{r})$ & $-\sqrt{2}\sin\zeta\cdot M_{\boldsymbol{Q}_{1}}^{({\rm ph})}(iB_{{\rm 2u}}^{(2)})$\tabularnewline
\cline{2-3} \cline{3-3} 
 & (c3) $\sin(\boldsymbol{Q}_{1}\cdot\boldsymbol{r})$ & $\sqrt{2}\sin\zeta\cdot M_{\boldsymbol{Q}_{1}}^{({\rm ph})}(iB_{{\rm 1g}}^{(1)})$\tabularnewline
\hline 
\multirow{2}{*}{$\tau_{z}(s_{x}-s_{y})$} & (d1) $\cos(\boldsymbol{Q}_{1}\cdot\boldsymbol{r})$ & 0\tabularnewline
\cline{2-3} \cline{3-3} 
 & (d3) $\sin(\boldsymbol{Q}_{1}\cdot\boldsymbol{r})$ & 0\tabularnewline
\hline 
\multirow{2}{*}{$\tau_{0}s_{z}$} & (e1) $\cos(\boldsymbol{Q}_{1}\cdot\boldsymbol{r})$ & $\frac{1}{\sqrt{2}}M_{\boldsymbol{Q}_{1}}^{({\rm ph})}(iB_{{\rm 1g}}^{(1)})-\frac{\cos\zeta}{\sqrt{2}}M_{\boldsymbol{Q}_{1}}^{({\rm ph})}(iB_{{\rm 1g}}^{(2)})$\tabularnewline
\cline{2-3} \cline{3-3} 
 & (e3) $\sin(\boldsymbol{Q}_{1}\cdot\boldsymbol{r})$ & $\frac{1}{\sqrt{2}}M_{\boldsymbol{Q}_{1}}^{({\rm ph})}(iB_{{\rm 2u}}^{(2)})+\frac{\cos\zeta}{\sqrt{2}}M_{\boldsymbol{Q}_{1}}^{({\rm ph})}(iB_{{\rm 2u}}^{(1)})$\tabularnewline
\hline 
\multirow{2}{*}{$\tau_{z}s_{z}$} & (f1) $\cos(\boldsymbol{Q}_{1}\cdot\boldsymbol{r})$ & $\frac{\sin\zeta}{\sqrt{2}}M_{\boldsymbol{Q}_{1}}^{({\rm ph})}(iA_{{\rm u}})$\tabularnewline
\cline{2-3} \cline{3-3} 
 & (f3) $\sin(\boldsymbol{Q}_{1}\cdot\boldsymbol{r})$ & $-\frac{\sin\zeta}{\sqrt{2}}M_{\boldsymbol{Q}_{1}}^{({\rm ph})}(iB_{{\rm 3g}})$\tabularnewline
\hline 
\multirow{2}{*}{$\tau_{0}s_{0}$} & $\cos(\boldsymbol{Q}_{1}\cdot\boldsymbol{r})$ & $\frac{1}{\sqrt{2}}M_{\boldsymbol{Q}_{1}}^{({\rm ph})}(A_{{\rm g}}^{(2)})+\frac{\cos\zeta}{\sqrt{2}}\cdot M_{\boldsymbol{Q}_{1}}^{({\rm ph})}(A_{{\rm g}}^{(1)})$\tabularnewline
\cline{2-3} \cline{3-3} 
 & $\sin(\boldsymbol{Q}_{1}\cdot\boldsymbol{r})$ & $\frac{1}{\sqrt{2}}M_{\boldsymbol{Q}_{1}}^{({\rm ph})}(B_{{\rm 3u}}^{(1)})+\frac{\cos\zeta}{\sqrt{2}}\cdot M_{\boldsymbol{Q}_{1}}^{({\rm ph})}(B_{{\rm 3u}}^{(2)})$\tabularnewline
\hline 
\multirow{2}{*}{$\tau_{z}s_{0}$} & $\cos(\boldsymbol{Q}_{1}\cdot\boldsymbol{r})$ & $\frac{\sin\zeta}{\sqrt{2}}M_{\boldsymbol{Q}_{1}}^{({\rm ph})}(B_{{\rm 1u}})$\tabularnewline
\cline{2-3} \cline{3-3} 
 & $\sin(\boldsymbol{Q}_{1}\cdot\boldsymbol{r})$ & $\frac{\sin\zeta}{\sqrt{2}}M_{\boldsymbol{Q}_{1}}^{({\rm ph})}(B_{{\rm 2g}})$\tabularnewline
\hline 
\end{tabular}

\caption{\label{tab:3}Projection of the density waves onto the low-energy
states at VHSs. Only the charge density waves and the spin density
waves with $\boldsymbol{Q}_{1}$ are listed. Charge current orders
and spin current orders are not considered. Each row corresponds to
a density wave $\Delta(\boldsymbol{r})$ taking the form $\sum_{\boldsymbol{R}}\sum_{a=A,B}M_{aa}\delta(\boldsymbol{r}_{\parallel}-\boldsymbol{R}-\boldsymbol{b}_{a,\parallel})\delta(z-b_{a,z})]$
multiplied by $\cos(\boldsymbol{Q}_{1}\cdot\boldsymbol{r})$ or $\sin(\boldsymbol{Q}_{1}\cdot\boldsymbol{r})$.}
\end{table}

\subsection{Symmetry analysis of the degeneracies of density-wave channels in the RG results}
\label{subsec:degeneracy}

Three types of degeneraices are seen in Fig. \ref{fig:S4} and Fig.
\ref{fig:S5}. The first type is the degeneracy between the pairs
of irreps: $B_{{\rm 3g/u}}\leftrightarrow A_{{\rm u/g}}$, $B_{{\rm 1g/u}}\leftrightarrow B_{2{\rm u/g}}$,
and $B_{{\rm 2g/u}}\leftrightarrow B_{1{\rm u/g}}$. This degeneracy can be understood as the consequence of the broken translation symmetries (${E|100}$,${E|010}$) or the following gauge symmetry when $0<k<\pi$: 
\[
(\psi_{\boldsymbol{K}_{1}+\boldsymbol{k}},\psi_{\boldsymbol{K}_{2}+\boldsymbol{k}},\psi_{\boldsymbol{K}_{3}+\boldsymbol{k}},\psi_{\boldsymbol{K}_{4}+\boldsymbol{k}})^{T}\rightarrow(e^{i\frac{\pi}{4}}\psi_{\boldsymbol{K}_{1}+\boldsymbol{k}},e^{i\frac{\pi}{4}}\psi_{\boldsymbol{K}_{2}+\boldsymbol{k}},e^{-i\frac{\pi}{4}}\psi_{\boldsymbol{K}_{3}+\boldsymbol{k}},e^{-i\frac{\pi}{4}}\psi_{\boldsymbol{K}_{4}+\boldsymbol{k}})^{T},
\]
which is inherited from the gauge transformation $\Theta(\boldsymbol{k})$
in Sec. \ref{subsec:derivation}. As this gauge transformation leaves
all the thirteen interactions invariant in the current patch model,
the degeneracy is secured over the RG iteration. Also, it is useful
to note that $\sin(\boldsymbol{Q}_{1}\cdot\boldsymbol{r})$ is a basis
function of $B_{{\rm 3u}}$ in $D_{{\rm 2h}}$.

The second type is the degeneracy occuring at $\sin^{2}\zeta=0$ at
which $\boldsymbol{\alpha}_{{\rm R}}|_{{\rm VHS}}=0$. In this case,
the spin rotation symmetry is restored. Therefore, as all the spin-triplet
Cooper pairings are degenerate at this point, all the TRSB (TRS) density
waves becomes degenerate.

The last degeneracy type occurs at $\sin^{2}\zeta=1$ at which $\varepsilon_{x,\boldsymbol{k}}|_{{\rm VHS}}=0$.
At this point. the bilayer system is decomposed into two independent
single layers since the bare interaction does not contain an inter-layer
interaction. The independence of two single layer extends the symmetry
group of the Hamiltonian. For example, a coordinate transformation
$E\oplus C_{2z}$ which leaves a layer invariant while rotates the
other layer $180^{\circ}$ becomes available. Note that $\tau_{0}s_{i}$
with $i=1,2,3$ are transformed into $\tau_{z}s_{i}$ by $E\oplus C_{2z}$.

\bibliography{ref}